\documentclass[prx,showpacs,twocolumn,superscriptaddress,aps]{revtex4-2}

\usepackage[utf8]{inputenc}
\usepackage[american,british]{babel}
\usepackage[T1]{fontenc}
\usepackage[pdftex]{graphicx}  
\usepackage{graphicx, xcolor}
\usepackage{dcolumn}
\usepackage{braket}
\usepackage{bm}
\usepackage{amsmath,amsthm,amssymb}
\usepackage{color}
\usepackage{verbatim}

\usepackage[colorlinks,linkcolor=blue,citecolor=blue,urlcolor=blue]{hyperref}

\newcommand{\ee}{\mathrm{e}}
\newcommand{\iu}{\mathrm{i}}  
\usepackage{dsfont}
\usepackage{physics}
\usepackage{soul}

\newcommand{\new}[1]{#1}

\begin{document}
\title{Experimental violations of Leggett-Garg's inequalities on a quantum computer}
\date{\today}

\author{Alessandro Santini}
\email{asantini@sissa.it}
\affiliation{SISSA, via Bonomea 265, 34136 Trieste, Italy}

\author{Vittorio Vitale}
\email{vvitale@sissa.it}
\affiliation{SISSA, via Bonomea 265, 34136 Trieste, Italy}
\affiliation{The Abdus Salam International Center for Theoretical Physics, Strada Costiera 11, 34151 Trieste, Italy}

\begin{abstract}
Leggett-Garg's inequalities predict sharp bounds for some classical correlation functions that address the quantum or classical nature of real-time evolutions. We experimentally observe the violations of these bounds on single- and multi-qubit systems, in different settings, exploiting the IBM Quantum platform.
In the multi-qubit case we introduce the Leggett-Garg-Bell's inequalities as an alternative to the previous ones.
Measuring these correlation functions, we find quantum error mitigation to be essential to spot inequalities violations.
Accessing only two qubit readouts, we assess Leggett-Garg-Bell's inequalities to emerge as the most efficient quantum coherence witnesses to be used for investigating quantum hardware\new{, among those introduced}.
Our analysis highlights the limits of nowadays quantum platforms, showing that the above-mentioned correlation functions deviate from theoretical prediction as the number of qubits and the depth of the circuit grow.

\end{abstract} 

\maketitle

\section{Introduction}\label{sec:Intro}
As the effort in building fault-tolerant quantum computers increases, the need for efficient benchmarking and characterization of these devices becomes more and more apparent. Several studies have been realized in the direction of gauging current Noisy Intermediate-Scale Quantum (NISQ) devices~\cite{Preskill_2018}: some of them focus on the analysis of the entanglement behaviour of these systems and ways of measuring it \cite{brydges2019probing,vitale2021symmetry}, others propose new efficient entanglement detectors to be used in experiments \cite{elben2020mixed,neven2021symmetry,yu2021optimal}.  In this work, we want to exploit Bell's-like inequalities in time, dubbed Leggett-Garg's inequalities (LGIs)~\cite{LGI,Leggett_2002}, to test the quantum coherence of a quantum hardware. The availability of the open platform `IBM Quantum' \cite{IBMQ_ref} gives us the opportunity of carrying out a rigorous study of its performance highlighting, at same time, the power of LGIs as witnesses of quantum coherence, applying them to real-world experiments. 

A plethora of experimental tests of LGIs, and similar conditions, have been already performed on two-level systems \cite{palacios2010experimental,knee2012violation,xu2011experimental} as well as more complicated experimental setups like photonic systems, phosphorus impurities in silicon, superconducting devices, and nuclear magnetic resonances \cite{goggin2011violation,Groen2013,Emary2014,Budroni2015,katiyar2017experimental,wang2017enhanced,Bose2018}.
However, while numerical computation of LGIs to assess the quantum coherence of an open quantum systems has been carried out in the past years \cite{Friedenberger2017,vitale2019assessing}, their investigation along the real time-dynamics of quantum hardware has rarely been considered as topic of study \cite{Huffman2017,ku2020experimental}. 
As IBM Quantum becomes more popular \cite{Solfanelli21Arxiv,Tacchino20AQT,Devitt2016} and its performance \new{improves}~\cite{Bravyi2021,Jurcevic_2021,eddins2021doubling}, we gauge it as the perfect remotely programmable playground for analysing the \new{prowess} of LGI in detecting quantum coherence~\cite{Alsina2016,mooney2019entanglement,wang201816}.

Understanding the frontier between \new{quantum and classical} mechanics and how the \new{latter arises from the former} are open problems and vivid topics of study. Following the flood of the studies about how macroscopic quantum coherence could be realized in laboratory, Leggett and Garg wrote Bell’s-like inequalities that test correlations of the same system measured at different times; in contrast with spatial Bell’s inequalities that put constrains on the correlations of spatially separated systems \cite{LGI,Leggett_2002}.
The starting point of the LGIs is the definition of macrorealism. This is contained in a small set of principles which have been phrased as follows: i) Macroscopic realism per se (MRPS): a macroscopic object \new{with two or more, macroscopically dinstinct, available states is,} at any given time, in
a definite one of those states.  ii) Non-invasive measurability (NIM): it is possible in principle to determine \new{in which of these states the system is}, without any effect on the state itself or on
the subsequent system dynamics.
When one of these assumptions is not satisfied, one may assert that the system cannot be described by a macrorealistic theory, i.e. a theory \new{which adheres} to our intuition of \new{how the macroscopic world} should behave.
Thus, assuming a quantum theory as the only alternative to a macrorealistic theory, LGIs can distinguish between classical and quantum systems.

In calculating LGIs, one considers \new{consecutive measurements} on a single system at different times and assumes NIM to hold. A heated argument which rises from this assumption is the so-called `clumsiness loophole'. This argument can be summarized as follows: observing a LGI violation, a skeptical  macrorealist  might always  appeal  to  hidden  invasiveness of the measurements to  explain the  violations, since it is impossible to conclusively demonstrate that a physical measurement is in fact non-invasive~ \cite{wilde2012addressing}.
The analogous loophole in Bell’s inequalities is the communication loophole~\cite{Brunner2014}, which can be solved assuming that the measurements are sufficiently separated in space, so that one measurement cannot influence the other. This is commonly known as the principle of locality~\cite{benenti2019principles}.
As an effort to go beyond the clumsiness loophole in LGIs, a hybrid version of the LGIs has been investigated~\cite{Dressel2014,white2016preserving,Thenabadu2020}.
These conditions, called Leggett-Garg-Bell's inequalities (LGBIs), extend LGIs as both temporal and spatial Bell's like inequalities~\cite{Dressel2014,white2016preserving,Thenabadu2020,thenabadu2021bipartite}. They assume the measurement of spatially separated parts of a system at different times so that the principle of NIM may be substituted by the principle of locality ~\cite{benenti2019principles}. 

In general, most of the experimental tests of the LGIs conducted so far suffer from the `clumsiness loophole'~\cite{wilde2012addressing} and it is still an open question whether loophole-free Leggett-Garg's protocols can be constructed.
In the only cases in which LGIs have been implemented on the IBM Quantum architecture~\cite{Huffman2017,ku2020experimental}, to the best of our knowledge, the authors circumvent the problem of NIM by use of an alternative witness. They introduce a condition that defines a `clumsy macrorealism' which embodies the possibility of modifications of the state of the system after a measurement and discuss its properties in several settings.

Also, in the case of LGBIs, for the principle of locality to hold, one may observe that the parts of the system which are measured should be infinitely distant or distant enough according to the Lieb-Robinson bound~\cite{lieb1972finite}, condition which may not be satisfied in the settings we are interested in.

In this work we will not discuss these technicalities in detail, nevertheless we believe our work may stimulate the search for generalizations of LGIs/LGBIs where \new{the NIM/locality} assumption can be relaxed.
For a more detailed treatment of the loopholes of both LGIs and LGBIs we encourage the reader to consult Refs.~\cite{emary2013leggett,ku2020experimental} and references therein.

Evidently, quantum hardware may benefit from quantum coherence witnesses, like LGIs and LGBIs, acting as benchmark of their proper functionality. At the same time, investigating theoretical bounds of correlation functions on experimental setups is quite interesting on its own, from a fundamental standpoint.
With this in mind, we first introduce LGIs and LGBIs; then, we discuss the prototypical example of a single qubit (spin-$1/2$), \new{evolving} under a unitary dynamics, comparing the experimental results against the theoretical ones. We measure the LGIs for one of the IBM Quantum transmons, which plays the role of a quantum qubit, evolving with its own dynamics, experimentally showing the effects of decoherence on the hardware. This allow us to roughly estimate the value of the coherence time $\langle T_2 \rangle$ of the transmon, in agreement with the nominal value for the processor used and the effective dynamics which describes it.
Moreover, we investigate multi-qubit systems, calculating LGIs and LGBIs in different settings.
We observe LGIs and LGBIs violations in several experimental setups addressing the quantum coherence of the hardware in the timescales investigated.
Finally, we apply LGBIs to a many-body example, namely the transverse field Ising model, to show that the performance of the hardware \new{worsens with an increasing depth of the circuit}. This result allows us to elaborate on the quantum coherence of IBM Quantum processors in case of multi-qubit computations and draw our conclusions.
Error mitigation is exploited during this work to improve the performance of the readouts of the IBM Quantum processors. We observe that such procedure is essential to witness most of inequalities violations we address, thus showing the limits of the hardware and the power of LGIs/LGBIs alike.

\section{Leggett-Garg's Inequalities}
Leggett-Garg's inequalities predict thresholds for classical correlation functions which can be violated if the system behaves according to quantum mechanics. Their definition \new{is based on the concept} of macrorealism that is encoded in a set of assumptions which a classical system must \new{stick to} \cite{Leggett_2002,emary2013leggett}.
Based on the assumptions introduced above, Leggett and Garg derived Bell’s-like inequalities
that any system behaving `classically' should obey. Violations of these
inequalities provide evidence of quantum behavior of a system if it is accepted that
the alternative to classical theories is quantum mechanics. In App.~\ref{app:Derivation}, we present a detailed derivation of the LGIs, according to~\cite{emary2013leggett}, for the sake of completeness. Here we introduce the basic details and show the inequalities we will use for the remainder of the work.

Let us define a classical dichotomic variable Q which can assume \new{values} $+1$ or $-1$: $Q(t_i)=Q_i$  stands for the measurement value of the observable at time $t_i$. We denote with $P_i(Q_i)$ the probability of obtaining the result $Q_i$. The correlation function $C_{ij}$ is written as
\begin{equation}
C_{ij}= \sum_{Q_{i},Q_{j}=\pm1}Q_iQ_jP_{ij}(Q_i,Q_j),
\end{equation}
where the subscripts of $P$ remind us of the times at which the measurements are performed.
Assuming the principle introduced before to hold, one can prove that
\begin{equation}
\begin{aligned}
K_3 = C_{12} +C_{23} -C_{13},& \quad   -3\leq K_3 \leq 1;\\
K'_3 = -C_{12} -C_{23} -C_{13},& \quad  -3\leq K'_3 \leq 1;\\
K^{\textrm{perm}}_3 = -C_{12} +C_{23} +C_{13},&  \quad -3\leq K^{\textrm{perm}}_3 \leq 1.\\
\end{aligned}\label{eq:LGI3_def_equations}
\end{equation} 
These are all the conditions on the possible different correlations functions one can derive at third-order, namely performing three measurements in time, $t_1<t_2<t_3$.

Even though LGIs are not necessary conditions to assess the quantum coherence of the time evolution~\cite{SSMajidy2019,SSMajidy2021}, they have been successfully utilized to address this problem in the study of open quantum systems \cite{Friedenberger2017,vitale2019assessing}.

\section{Leggett-Garg-Bell's Inequalities}\label{sec:LGBI}
Leggett-Garg's inequalities can be computed on any kind of systems as the only prerequisite is the definition of a dichotomic variable. Taking into account a large ensemble of qubits it is possibile to exploit them considering a global variable, e.g. total angular momentum or spin, as done in Ref.~\cite{Budroni2014,Lambert2016}. From an experimental point of view, it may lead to bigger errors in the computation as it implies the readout of all the qubits.
Remarkably, it is possible to take into account different inequalities that circumvent this problem.
The rationale behind this extension is to substitute the NIM postulate with the one of locality as anticipated before.
In this framework, measurements are made at different times and at distinct locations,  so  that  locality can be  invoked to  justify non-invasiveness.
The analytical derivation of the inequalities is equivalent to the one of the LGIs where the second measurement is performed on a different location. 
Here we define the correlation function $C^{AB}_{ij}$ as 
\begin{equation}
C^{AB}_{ij}= \sum_{Q^A_{i},Q^B_{j}=\pm1}Q^A_iQ^B_jP^{AB}_{ij}(Q^A_i,Q^B_j),
\end{equation}
where the subscripts $i,j$ denote the times when the measurements are performed and the superscripts $A,B$ are two spatially-separated locations.
Considering macroscopic realism and `Bell/macroscopic locality' one gets~\cite{Thenabadu2020,thenabadu2021bipartite}
\begin{equation}
\begin{aligned}
T_3 = C^{AB}_{12} +C^{AB}_{23} -C^{AB}_{13},& \; \;  -3\leq T_3 \leq 1;\\
T'_3 = -C^{AB}_{12} -C^{AB}_{23} -C^{AB}_{13},& \; \; -3\leq T'_3 \leq 1;\\
T^{\textrm{perm}}_3 = -C^{AB}_{12} +C^{AB}_{23} +C^{AB}_{13},&  \; \; -3\leq T^{\textrm{perm}}_3 \leq 1.
\end{aligned}\label{eq:LGBI3_def_equations}
\end{equation} 
at third-order, i.e. considering three time instants.

The introduction of the LGB's correlation functions $T$ will be very useful for us in the context of gauging IBM Quantum.
In fact, since they involve the measurement of only two qubits but intrinsically take into account the spatial degrees of freedom of the system, they might be a valid alternative to LGIs on multi-qubit systems.

\section{Single qubit experiments}
\subsection{Time evolution of a single qubit}
\begin{figure}
    \centering
	\includegraphics[width=\linewidth]{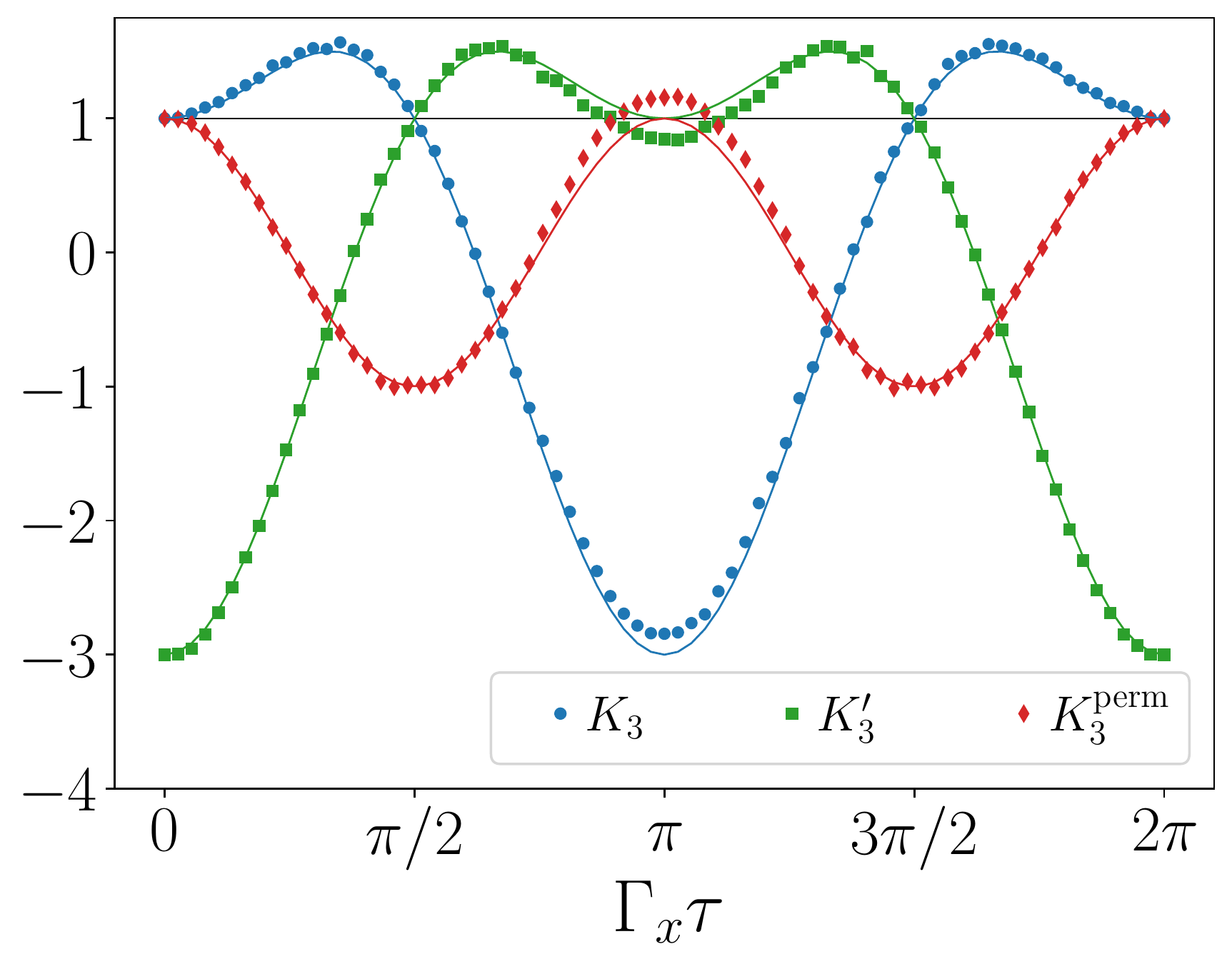}
	\includegraphics[width=0.9\linewidth]{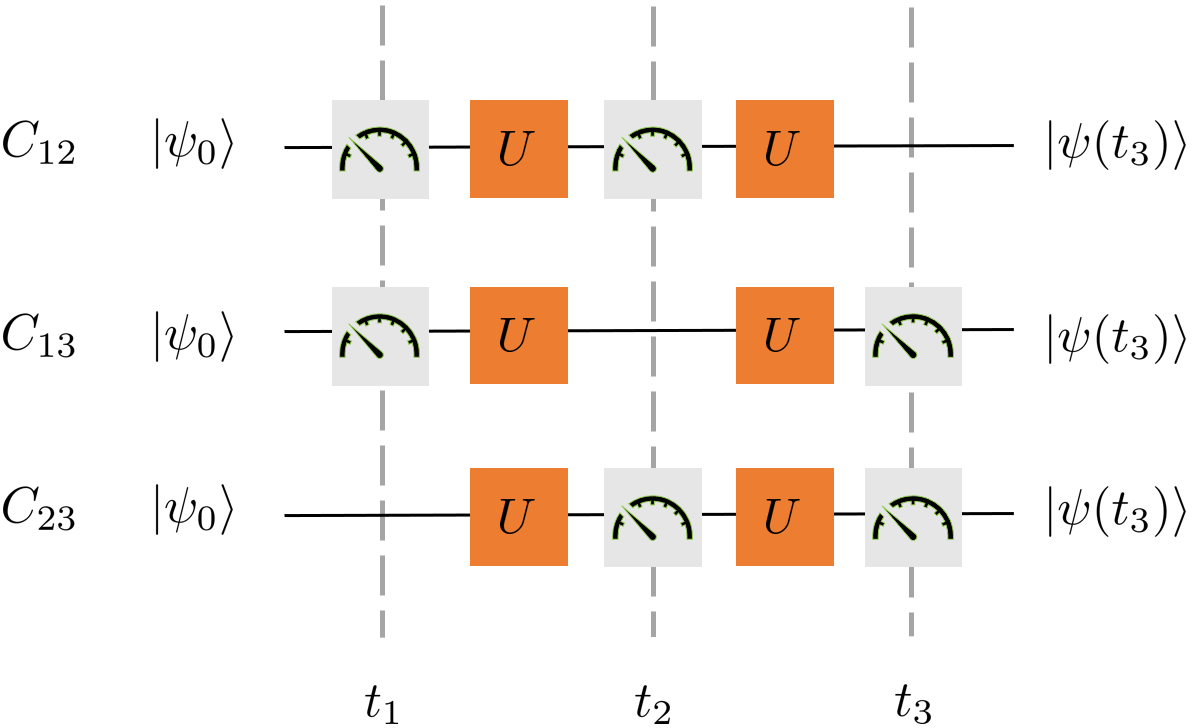}
	\caption{	Top: LGIs as a function of the time difference between two subsequent measurements (circuit depth $7$). The curves show  $K_{3}$, $K'_3$ and $K^{\textrm{perm}}_{3}$, according to Eq.~(\ref{eq:LGI}). The violation threshold is marked by the solid black line. Markers are experimental results, solid lines are theoretical predictions. \new{The error bars are not visible as the statistical error is about $10^{-2}$.}
	Bottom: Diagram of the evolution and the measurement scheme.}\label{fig:LGI_qubit}
\end{figure}
The starting point of the investigation of \new{LGIs} would be naturally the canonical example of a qubit evolving under the \new{Hamiltonian}
\begin{equation}\label{eq:QubitH}
H=\frac{\Gamma _x \sigma _x}{2},
\end{equation} 
where \new{$\Gamma_x$ is the qubit frequency (we set $\hbar=1$)}. It can be used both as an introduction to how \new{LGIs} work and can be interpreted, as well as a first benchmark of the IBM Quantum hardware. The analytical calculation \new{of the LGIs} for a two-level system is straightforward and proceeds as follows. First, \new{we choose $\hat{Q}=\sigma_z$ as the dichotomic operator taking values} $\pm1$ if the z-component of the spin of the qubit is up/down; secondly, since the evolution operator
\begin{equation}U(t) = e^{-i\frac{\Gamma _x \sigma _x}{2}t},
\end{equation}
is a simple rotation around the x-axis, the correlation $C_{ij}$ takes the analytical expression \cite{emary2013leggett,fritz2010quantum}:
\begin{equation}
C_{ij}=\cos{\Gamma_x(t_j -t_i)}.
\end{equation}
If we set $t_3-t_2=t_2-t_1=\tau$ we can express the functions $K_3,K'_3,K^{\textrm{perm}}_{3}$ in terms of the time difference between the two measurements:
\begin{equation}\label{eq:LGI}
\begin{aligned}
&K_3=2\cos{\Gamma_x\tau}-\cos{2\Gamma_x\tau},\\
&K'_3=-2\cos{\Gamma_x\tau}-\cos{2\Gamma_x\tau},\\
&K^{\textrm{perm}}_{3}=\cos{2\Gamma_x\tau}.
\end{aligned}
\end{equation}
These functions oscillate in time and violate the inequalities only for certain values of  $\tau$.
The quantities in Eq.~(\ref{eq:LGI}) are plotted in Fig.~\ref{fig:LGI_qubit} as continuous curves.
As shown in Fig.~\ref{fig:LGI_qubit} and observed in Ref.~\cite{huelga1995proposed},  $K_3$ and $K'_3$ appear to be complementary: one is violated when the other is not and vice-versa.
\new{This complementary behavior allows a detection of the non-classical properties of the two-level system over the full parameter range.}\\
It can be seen that this is just a particular case due to the specific choice of the Hamiltonian and the dichotomic variable. Hence one should not be \new{misled} that third-order LGIs can provide a comprehensive picture of the physics of the system.
However, as already pointed out, the comparison between theoretical and experimental calculation of LGIs may be informative of the performance of the hardware and the used experimental setup. 
\\
The evolution of the system considered in this section is straightforwardly implemented in IBM Quantum. We consider $N_{\tau}=75$ values for $\tau \in [0,2\pi/\Gamma_x]$ and evaluate $C_{ij}$, in particular 
\begin{equation}
\begin{aligned}
    &C_{12}(\tau)=\langle \sigma^{z}(0) \sigma^z(\tau)\rangle,\\
    &C_{13}(\tau)=\langle \sigma^{z}(0) \sigma^z(2\tau)\rangle,\\
    &C_{23}(\tau)=\langle \sigma^{z}(\tau) \sigma^z(2\tau)\rangle,
    \end{aligned}\label{eq:C_ij_single_qubit}
\end{equation}
measuring the spin along the z-axis at the two appropriate time instants as shown in the bottom panel of Fig.~\ref{fig:LGI_qubit}. We repeat each evolution, together with the needed measurements, $n_{\textrm{shots}}=2^{13}$ times.
In the upper panel we show the results of the experiment performed on the IBM Quantum open access hardware `ibmq\_manila' (markers) and the analytical predictions according to Eq.~(\ref{eq:C_ij_single_qubit}) (solid lines). Error bars are not visible as the error on the measurement outcomes should be estimated of order $1/\sqrt{n_\mathrm{shots}}\approx 10^{-2}$. 
\\
\new{From Fig.~\ref{fig:LGI_qubit} is evident that the experimental results deviate from Eq.~(\ref{eq:LGI}) around $\Gamma_x\tau \sim \pi$ even though the qualitative picture is the same. 
Remarkably, while the results are not perfectly in agreement with the theoretical ones, we see that the LGIs threshold is always violated, witnessing that the interaction with the environment does not spoil the quantum dynamics.}
We seize this opportunity to mention that all the shown experimental results are obtained performing error mitigation. A detailed explanation on how it works and its huge influence on the outcomes is given in App.~\ref{app:ErrorMitigation}.

\subsection{Transmon qubit}
\begin{figure}
    \centering
    \includegraphics[width=\linewidth]{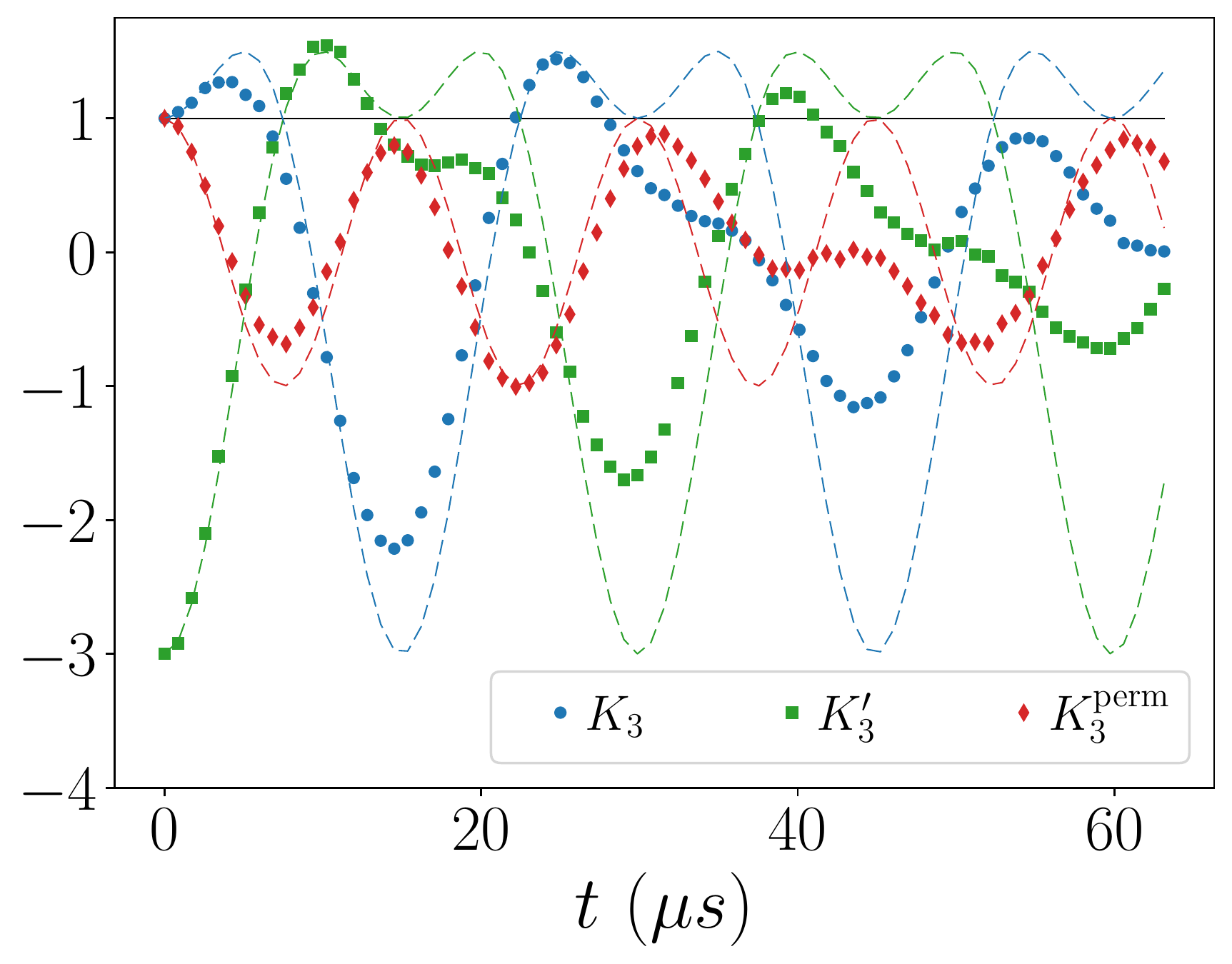}
    \caption{LGI as a function of the time difference between two subsequent measurements calculated on the transmon qubit evolution. The dashed curves are the theoretical predictions according to Eq.~(\ref{eq:LGI}), the markers are the experimental results (statistical error is less than the size of the markers). \new{A violation does not occur after $t \sim 45 \mu s$ in the experimental data, witnessing the loss of coherence of the system}. This is compatible with IBM Quantum estimation (at the time of the experiment) of the coherence time for the used system (`ibmq\_manila'): $\langle T_2 \rangle =55 \mu s$. \new{The error bars are not visible as the statistical error is about $10^{-2}$.}}
    \label{fig:transmon}
\end{figure}

An interesting case study, in relation with IBM Quantum, is analysing the evolution of one of the qubits of the quantum hardware when it is left untouched by the quantum circuit. The processors provided by IBM Quantum are constituted by transmons which can be modeled, for our purposes, as a two-level system described by the Hamiltonian:
\begin{equation}
    H^{\textrm{eff}}= - \frac{\hbar\Omega }{2}\sigma^z.
\end{equation}
It is possible to measure the LGIs on this system in the following way. Since one is interested in the dynamics of the transmon according to $H^{\textrm{eff}}$, one can force the hardware to evolve acting on an ancilla qubit with identity gates. In this way the transmon we are interested in would not be touched by the quantum gates and would evolve under its intrinsic `physical' Hamiltonian $H^{\textrm{eff}}$.
In $H^{\textrm{eff}}$, $\Omega \simeq 4.971 \;\mathrm{GHz}$ is the nominal frequency of the qubit used for the experiment~\cite{IBMQ_ref}.
Since the default initial state of the processors is $\ket{0}$, the qubit is rotated in the state $\ket{+}=\frac{1}{\sqrt{2}}(\ket{0}+\ket{1})$ by use of an Hadamard gate at the initial time. Then the evolution is executed.
The qubit rotates in the same fashion discussed in relation to Fig.~\ref{fig:LGI_qubit}, with its own intrinsic frequency $\Omega$.
In Fig.~\ref{fig:transmon} we plot the LG's correlation functions $K$ as a function of time. We observe that the threshold $1$ is violated multiple times, for $t<30\mu s$. For long \new{evolution times,} the dynamics is damped witnessing the loss of coherence in the system. 
The average coherence time estimated for the quantum hardware we used (`ibmq\_manila'), according to IBM, is $\langle T_2 \rangle \sim 55\mu s$.
We observe that assuming the theoretical behaviour to hold, i.e. without any damping (dashed lines), $K_3$ should have exceeded the threshold periodically. However, the \new{the experimental results do not violate the LGIs where one would have expected theoretically ( $t\sim 45 \mu s$)}. \new{This is} quite in agreement with the coherence time $\langle T_2 \rangle\sim55 \mu s$ predicted by IBM Quantum.
Here we point out that the nominal coherence time average $\langle T_2 \rangle$ is estimated at each recalibration of the processors. The value we write here is the one reported at the time the experiment was performed.
\begin{figure*}
    \begin{minipage}{\linewidth}
    \includegraphics[width=0.32\linewidth]{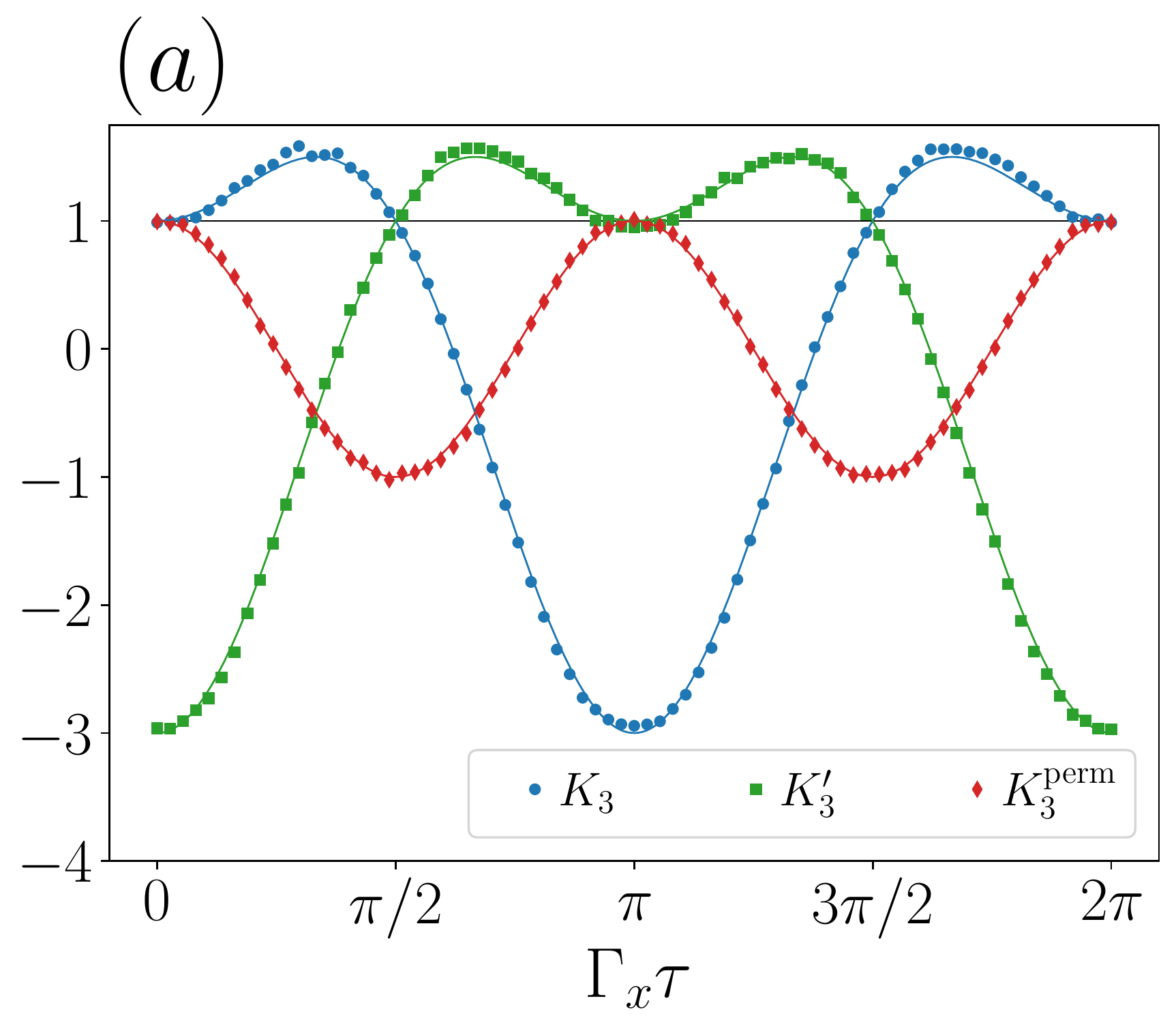}
    \includegraphics[width=0.32\linewidth]{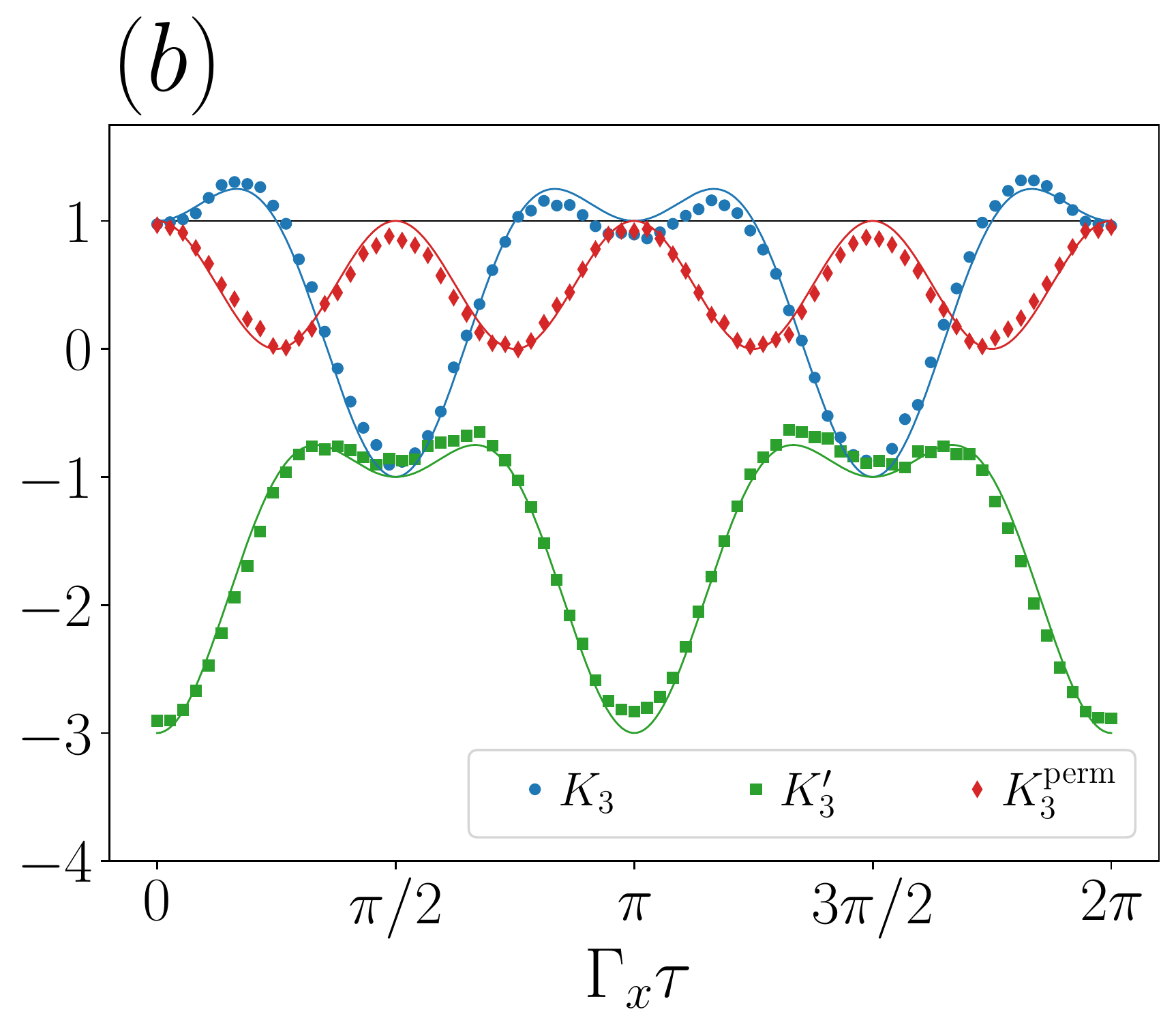}
    \includegraphics[width=0.32\linewidth]{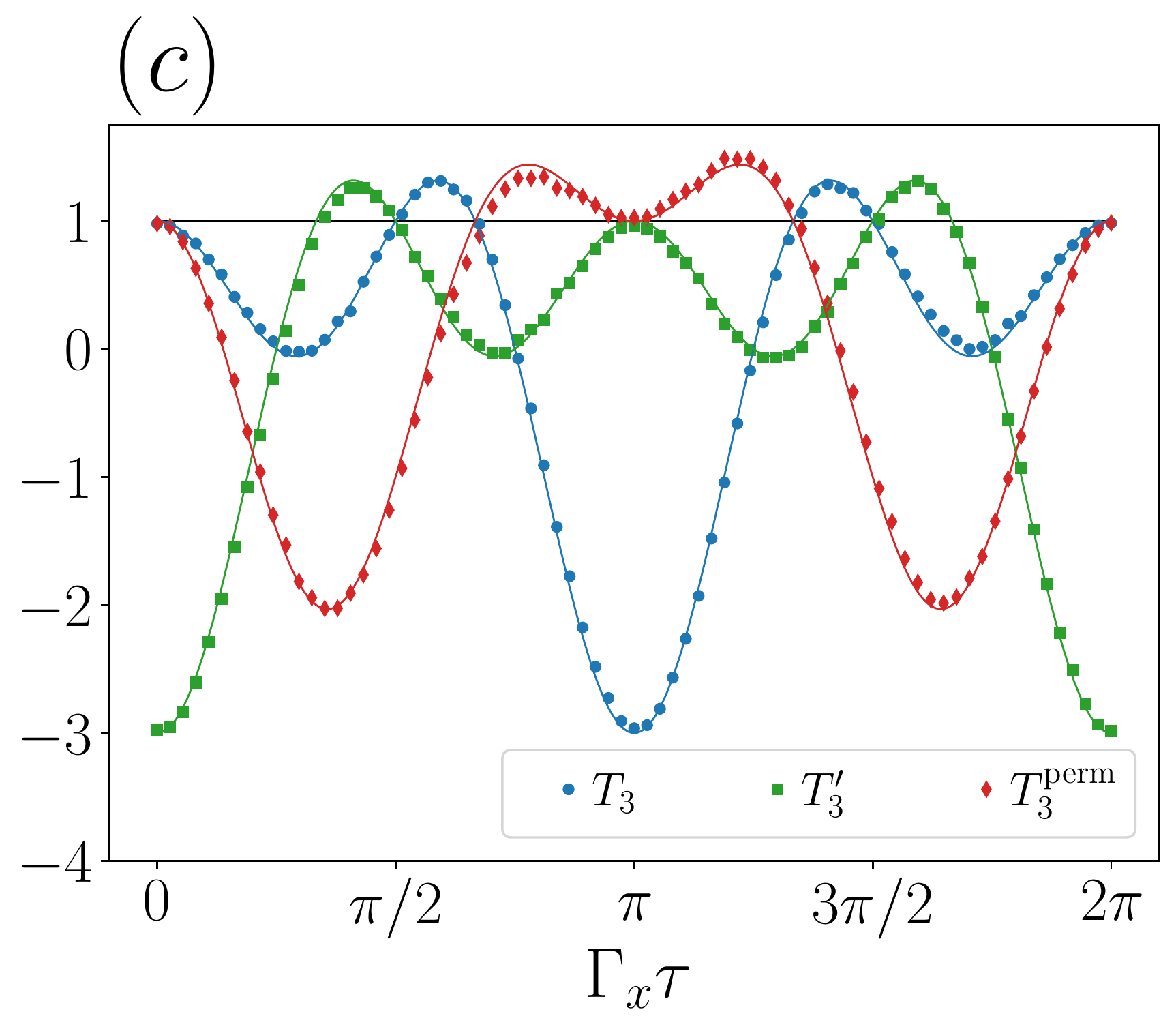}
    \end{minipage}
    \begin{minipage}{\linewidth}\centering
    \includegraphics[width=0.55\linewidth]{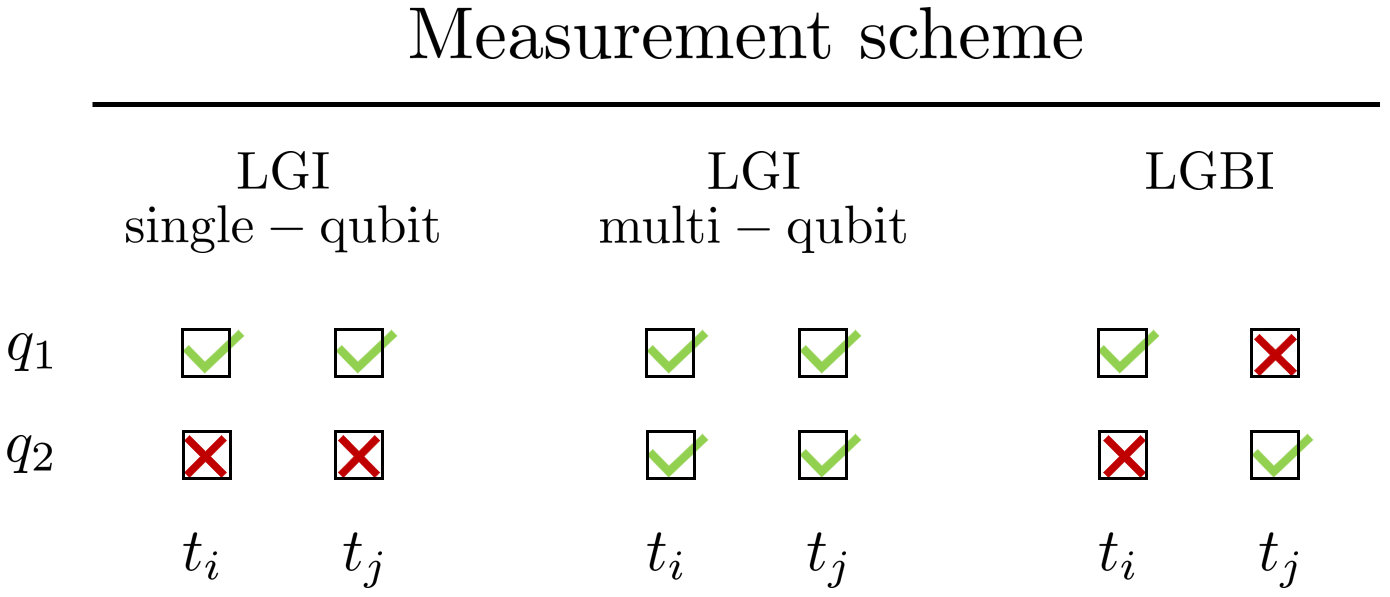}\hspace{.7cm}\includegraphics[width=0.4\linewidth]{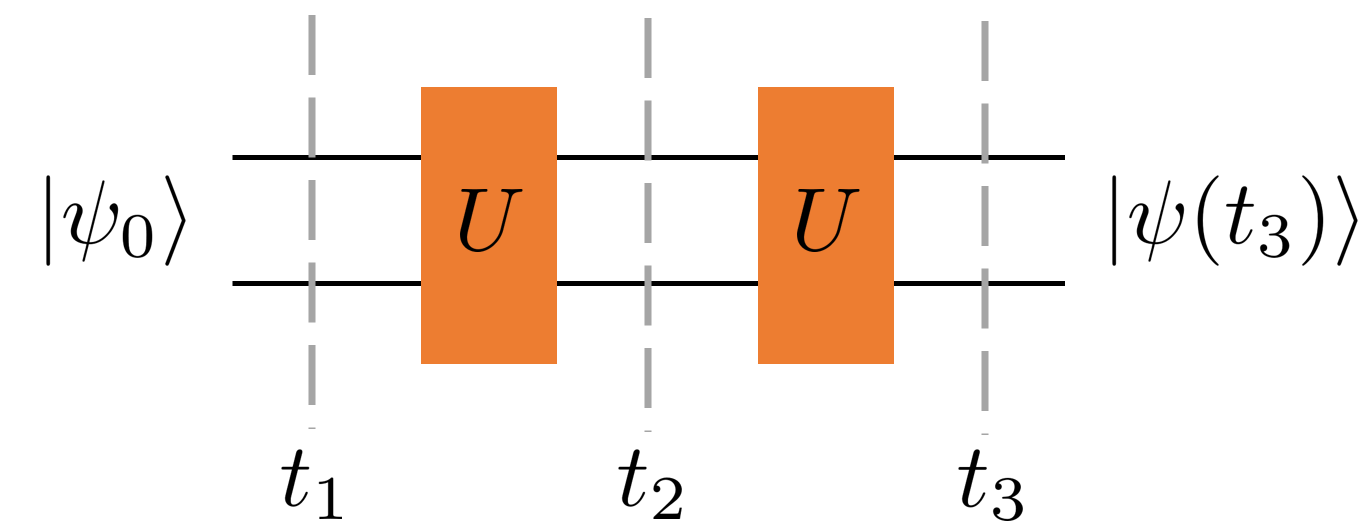}
    \end{minipage}
    \caption{LGBIs and LGIs on a two-qubits system. Markers \new{are} experimental results (statistical error is less than the size of the markers), solid lines \new{are} numerical predictions \new{obtained with exact diagonalization}. In panel $(a)$ LGIs calculated on a single qubit (circuit depth $11$); in panel $(b)$ LGI sconsidering the total spin of the system (circuit depth $11$); in panel $(c)$ LGBIs (circuit depth $10$). The horizontal line marks the threshold of the inequalities. Bottom: \new{Measurement scheme for calculating a LGI/LGBI and diagram of the evolution}. Red crosses \new{in the measurement scheme} label qubits which are not measured at times $t_i$ and $t_j$, green checkmarks denote the qubits that are measured. }
    \label{fig:LGB_bellstate}
\end{figure*}

\section{Multi-qubit experiments}\label{sec:multiqubit}
The ultimate goal of quantum computation is realizing a scalable and fault-tolerant quantum computer, which allows the usage of a large ensemble of qubits, taming the errors \new{accumulating during} the computation.
For this reason we want to investigate the behaviour of the \new{inequalities} we have introduced in multi-qubit systems, to highlight their usefulness and efficiency. 
In multi-qubit systems the LGIs would require a dichotomic variable which is defined on the entire ensemble \cite{Lambert2016}. For instance, one could choose the total spin along the z-axis $S^z=\sum_i\sigma^z_i$.
With increasing system size, however, not only the sources of error during the computation accumulates, but also the ones due to readouts, when measuring the whole ensemble of qubits.
We point out that the readout error is a huge constraint for IBM Quantum and error mitigation complexity scales exponentially with the system size. \new{For this reason one would like to narrow the number of qubits to be measured to the smallest possible amount. With this in mind}, \new{we will exploit LGBIs} which, by definition, take into account both temporal and spatial correlations in the system and only require two readouts. We will compare \new{the LGBIs} with the LGIs, \new{that can be defined} both performing \new{measurements} on a single qubit and \new{on} all the qubits, and discuss their limitations.

\subsection{Non-interacting qubits}
Let us start by considering a two-qubits system. We investigate LGIs calculated performing measurements on a single qubit (LGIs single-qubit), LGI where the dichotomic variable is a global one (LGIs multi-qubit), i.e. measuring all the qubits,  and LGBIs where two readouts on two different qubits are realized.
\new{We examine the dynamics of two qubits}, initially entangled, evolving independently according to the Hamiltonian
\begin{equation}\label{eq:NoninteractingH}
H=\sum_{i=1}^N \frac{\new{\Gamma}_i}{2} \sigma^x_i,
\end{equation}
and measured at times $t_1,t_2,t_3$; here $\new{\Gamma}_i=\Gamma_x\; \forall i$. 
\new{As before, we will assume} $t_2-t_1=t_3-t_2=\tau$ and plot the results as a function of the time difference $\tau$.
The initial state of the system is a Bell state 
\begin{equation}
    \ket{\psi}=\frac{1}{\sqrt{2}}(\ket{00}+\ket{11}),
\end{equation}
obtained by use of an Hadamard gate and a CNOT.
For each setup we have introduced above (LGIs single-qubit, LGIs multi-qubit, LGBIs), we take $N_{\tau}=75$ values for $\tau \in [0,2\pi/\Gamma_x]$ and repeat each simulation $n_\textrm{shots}=2^{13}$ times on `ibmq\_manila'.
The continuous lines in Fig.~\ref{fig:LGB_bellstate} are obtained calculating the correlation function exactly according to
\begin{widetext}
\begin{equation}\label{eq:analyticalCij}
    C_{ij}(t_i,t_j)=\langle \hat{Q}(t_i) \hat{Q}(t_j) \rangle =\sum_{n,m} q_n q_m\Tr{\Pi_m U(t_j,t_i) \Pi_n U(t_i,0) \rho_0 U^{\dagger}(t_i,0) \Pi_n U^{\dagger}(t_j,t_i) \Pi_m}.
\end{equation}
\end{widetext}
Here $q_n$ are the possible values of the dichotomic observable $\hat{Q}$, $\Pi_n$ are the corresponding projection operators, $U(t_j,t_i)=\ee^{-\iu H t}$ is the evolution operator and $\rho_0$ is the density matrix of the system at time $t=0$. This expression can be easily calculated analytically in case of non-interacting Hamiltonians~\cite{Lambert2016} and numerically in case of interacting-ones. 
A sketch of the evolution and of the measurement scheme is depicted in the bottom panels of Fig.~\ref{fig:LGB_bellstate}. Each upper panel plot corresponds to a different setup described in ``Measurement scheme'': the rows $q_1,q_2$ correspond to the two qubits of the system and the columns $t_i,t_j$ denote the times at which measurements are performed. We mark the square box with a red cross if a qubits is left untouched by the measurements at a given time; we check it with green checkmarks if the qubit is measured instead.

In panel $(a)$ of Fig.~\ref{fig:LGB_bellstate} we perform two measurements, for each choice of $\tau$, on the same qubit (LGIs single-qubit). We choose as dichotomic variable $\sigma^z_1$ (considering $\sigma^z_2$ would be analogous) and calculate the LGIs. Evidently the result is equivalent to the one in Fig.~\ref{fig:LGI_qubit}. We ascribe this behaviour to the fact that \new{single-qubit LGIs} are measuring the coherence of the single qubit and they cannot provide information on the full system, even if the qubits are entangled at the beginning of the evolution. 
In panel $(b)$ of Fig.~\ref{fig:LGB_bellstate} we calculate the LGIs choosing a dichotomic observable defined on the whole system, namely $\hat{Q}=2\left|\sum_i \sigma^z_i\right|-1=\sigma^z_1\sigma^z_2$ (LGIs multi-qubit). Evidently a violation of the LGIs occurs. This allows us to affirm that LGIs can be used to detect the quantum coherence during the evolution in the time interval considered. Since we are using a global observable, we are confident that the information provided concerns the whole system. The bottleneck of \new{multi-qubit LGIs} is the \new{required} number of single-qubit readouts. Since it scales with the size of the system and error mitigation is mandatory, as discussed in App.~\ref{app:ErrorMitigation}, it is not affordable for large system sizes. 

Finally in panel $(c)$ of Fig.~\ref{fig:LGB_bellstate} we calculate the LGBIs measuring $\sigma^z_1$ and $\sigma^z_2$.
We observe that also LGBIs are able to detect the quantum coherence of the evolution. In this experimental set-up, the violation is much stronger than in the case of LGIs multi-qubit, however one cannot assert with generality that LGBIs exhibits stronger violations with respect to LGIs in all cases.\\
\new{Let us comment that LGBIs can detect the quantum coherence with only two single-qubit readouts instead of measuring an observable defined on the whole system. Evidently, it would be interesting to expand the investigation to larger systems. For this reason in the following we will focus on LGBIs, to observe if they are able to witness the quantum coherence of the system also in a many-body scenario.} 
\\

\begin{figure*}
    \includegraphics[width=\linewidth]{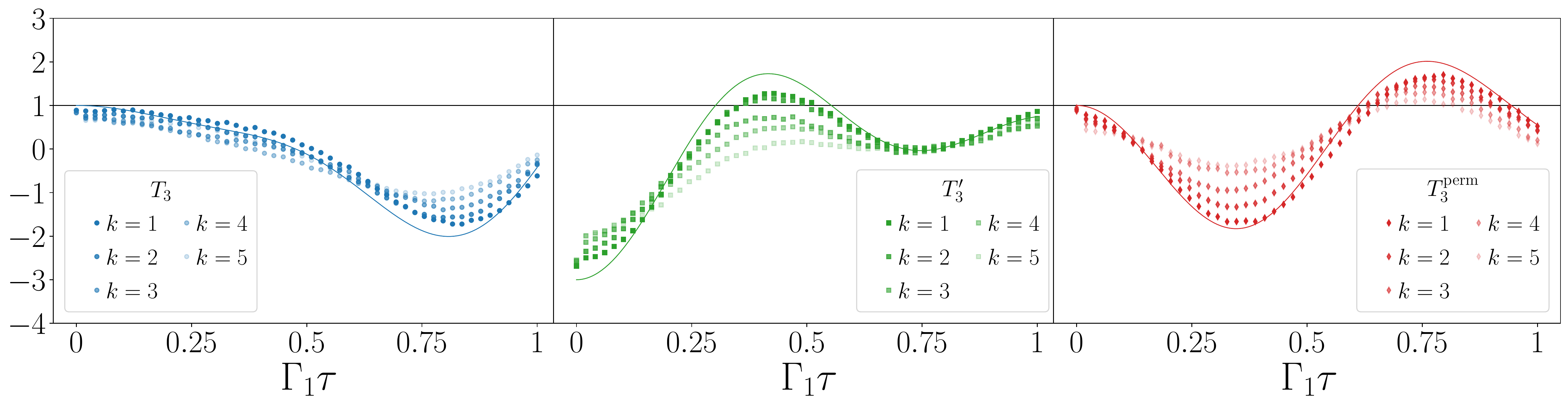}
    \caption{Transverse field Ising Chain, $J=0.1$ $\new{\Gamma}_i=1$, i=1,..,4, $\new{\Gamma}_5=2$. Markers are experimental results (statistical error is less than the size of the markers); the curves are obtained according to Eq.~(\ref{eq:analyticalCij}). We observe an experimental violation of the LGBIs.}
    \label{fig:TFIC04}
\end{figure*}

\subsection{Interacting qubits}

\new{Having} understood that the family of inequalities we have introduced can be successfully exploited to witness the quantum coherence of a quantum evolution,
from now \new{on} the focus of the study will be the performance of the hardware used, investigated exploiting LGBIs.
In particular, we will discuss an interacting quantum many-body system to elaborate on the performances of IBM Quantum in a physically intriguing setup. 

Simulating interacting systems on quantum computers is well-known to be a \new{famously daunting task} and bound \new{by} very small coherence times \cite{cervera2018exact,smith2019simulating}.
In this respect, we decided to consider an interacting evolution where the interaction contribution is a slight perturbation over a non-interacting dynamics.
 
Here we investigate a system of $N=5$ qubits, initially entangled in a GHZ state $(\ket{00000} + \ket{111111})/\sqrt{2}$, to be reproduced on `ibmq\_manila', during a transverse-field-Ising-chain-like (TFIC) evolution described by the Hamiltonian
\begin{equation}
H = -J\sum_{i=1}^{N-1} \sigma^z_i\sigma^z_{i+1} - \sum_{i=1}^N \new{\Gamma}_i \sigma^x_i;
\end{equation}
$N=5$ is the maximum number of qubits available in `ibmq\_manila'. We measure $\sigma_1^z$ and $\sigma_5^z$ to calculate the LGBIs. We consider $J=0.1$ and $\new{\Gamma}_i=1$ for $i=1,\dots,4$, $\new{\Gamma}_5=2$. The choice of the parameters is made in order to have a violation of the inequalities for $\new{\Gamma}_1\tau < 1$ \new{(see App.\ref{app:EPIE} for more details)}.

In order to implement the evolution on the quantum hardware, exploiting the fact that the interaction is nearest-neighbors, we split the Hamiltonian in $H_\mathrm{even} = \sum\limits_{i\ \mathrm{even}}\sigma^z_i\sigma^z_{i+1}$ and $H_\mathrm{odd} = \sum\limits_{i\ \mathrm{odd}}\sigma^z_i\sigma^z_{i+1}$, then we approximate $U(dt) \approx e^{-iH_\mathrm{even}dt}e^{-iH_\mathrm{odd}dt}$.
\new{To investigate the behaviour of IBMQ for different circuit depths, we set the number of applications of $U(dt)$ at $k=1,2,3,4,5$ with $dt = \tau/k$.}

In Fig.~\ref{fig:TFIC04} the experimental outcomes of \new{$T_3,T_3^{'},T_3^{\textrm{perm}}$} are shown together with the theoretical predictions. 
\new{In each panel we plot several curves for different values of the number of Trotter steps, as a function of time. In Table \ref{tab:depths} we write the circuits depths for the different implementations considered.
While the agreement between numerical and experimental simulations should improve increasing the number of steps in the Trotterization, as the approximation of the evolution is more accurate, we observe that the experimental results deviate more and more from the theoretical prediction with increasing $k$. We believe this is due to the errors accumulating during the computation, which grow with the number of gates implemented in the circuit.
Furthermore, we observe that while the agreement between experiments and numerical simulations is not perfect, the qualitative behaviour is somehow reproduced and it is possible to observe a violation of the inequalities both in $T_3^{'}$ and $T_3^{\textrm{perm}}$ for several values of circuit depths. 
This could be used to estimate the coherence time of the many-body system, since considering longer simulations one could expect that no other violation of the inequalities occurs as the number of gates applied increases.} 



\begin{table}
    \begin{tabular}{|r|r|r|}\hline
    Trotter steps&$C_{12}$ (depth)&$C_{13},\ C_{23}$ (depth)\\
    \hline
    \hspace{1cm}$k=1$\hspace{1cm}&\hspace{1cm}$18$\hspace{1cm}&\hspace{1cm}$29$\hspace{1cm}\\
    \hspace{1cm}$k=2$\hspace{1cm}&\hspace{1cm}$29$\hspace{1cm}&\hspace{1cm} $51$\hspace{1cm}\\
    \hspace{1cm}$k=3$\hspace{1cm}&\hspace{1cm}$40$\hspace{1cm}&\hspace{1cm} $73$\hspace{1cm}\\
    \hspace{1cm}$k=4$\hspace{1cm}&\hspace{1cm}$51$\hspace{1cm}&\hspace{1cm}$95$\hspace{1cm}\\
    \hspace{1cm}$k=5$\hspace{1cm}&\hspace{1cm}$62$\hspace{1cm}&\hspace{1cm}$117$\hspace{1cm}\\
    \hline
\end{tabular}
\caption{\new{Circuit depth of the circuits related to Fig. \ref{fig:TFIC04} for different values of the number of Trotter steps. In column $C_{12}$ and $C_{13},C_{23}$ we write the circuit depths related to the circuits used to estimate $C_{12}$ and $C_{13},C_{23}$ respectively. }}\label{tab:depths}
\end{table}
\section{Conclusions}
With the advent of modern technologies, the study of entanglement and quantum coherence of physical systems is crossing the boundaries of purely theoretical interest and is starting to intertwine with the practical investigation of now available quantum platforms. For this reason, reliable entanglement detectors and entanglement witnesses are becoming more and more important.
In this work we have decided to exploit IBM Quantum processors to investigate a class of well-known quantum coherence witnesses.
We have introduced Legget-Garg's inequalities (LGIs) and Leggett-Garg-Bell's inequalities (LGBIs) and measured their outcomes in different physical set-ups, taking advantage of the versatile programmable nature of IBM Quantum. 

We have shown experimental violations of the LGIs and LGBIs in single- and multi-qubit systems. We have observed that, investigating reasonable timescales, the LGIs are in agreement with theoretical results in non-interacting systems, witnessing that the IBM Quantum processor used (`ibmq\_manila') is robust against decoherence at early times.
Furthermore, we have explicitly investigated the coherence of one of the transmons constituting `ibmq\_manila' processor. We let it evolve under its own dynamics, according to the effective transmon Hamiltonian, and observed that LGIs violations do occur up to a characteristic time which is compatible with the nominal coherence time of the system $\langle T_2 \rangle$. The nominal value of $\langle T_2 \rangle$ considered is the one which has been estimated at the time of the experiment.

To further deepen the investigation of the Legget-Garg's-like inequalities we have considered two-qubits systems where the qubits, initially entangled in a Bell's state, evolve independently. We have made a comparison among LGIs using a single qubit dichotomic variable, LGIs using a global observable and LGBIs.
We have shown that LGBIs provide the same violation of the threshold as LGIs in the time range investigated requiring only two readouts. As a consequence one can focus on the calculation of LGBIs for studying multi-qubit systems.

Furthermore we have elaborated on the necessity of error mitigation for the observation of the inequalities violations in a wide range of the time interval. Error mitigation complexity scales exponentially with the system size and becomes quickly unfeasible for the estimation of LGIs using multi-qubits observable. Then, LGBIs show also the merit of requiring only two measurements whatever the size of the system, such that error mitigation is always possible after the experiments.

To apply the condition we have introduced to a relevant physical case and \new{to} benchmark the IBM Quantum platform, we discussed a quantum many-body problem where interaction is added as a small perturbation on a non-interacting dynamics. \new{In this simple case, where the value of the parameters is fine-tuned, we observed experimental results violate the threshold, but deviating from the curves predicted by analytical results. Furthermore we showed that quantum coherence is lost increasing the number of number of gates of the system.}

In conclusion, our results both show the experimental observation on a quantum computer of violations of Leggett-Garg's type inequalities and provide evidence of the usefulness of these quantum witnesses in the context of NISQ devices.
\new{The results also suggest that, increasing the circuit depth, even LGBIs, which emerge as the most efficient witnesses, among the ones introduced, rapidly depart from the theoretical prediction and may fail in detecting quantum coherence, since the gates errors accumulate.}

As we have observed that Leggett-Garg's type inequalities can be successfully used to discriminate against classical or quantum evolution, one would be interested in checking their behaviour with larger system sizes and more complex quantum circuits.
It could \new{also be} interesting to investigate particular topologies, not \new{currently} available in IBM Quantum, which allow to connect an ancilla to all the qubits performing a given computation, in order to assess if ancilla-based methods may be beneficial in this framework.\\

\section*{Data availability}
The data that support the plots within this paper and other findings of this study are available from the authors upon request.

\section*{Competing interests}
The authors have no potential financial or non-financial conflicts of interest.\\

\section*{Acknowledgments}
We are grateful to M. Dalmonte, G. Santoro \new{and A. Galvani} for valuable discussions.

We acknowledge the use of IBM Quantum services for this work \cite{IBMQ_ref}. The views expressed are those of the authors, and do not reflect the official policy or position of IBM or the IBM Quantum team. 

VV acknowledges support by the ERC under grant number 758329(AGEnTh), has received funding from the European Union’s Horizon 2020 research and innovation programme under grant agreement No 817482 (Pasquans).

The authors acknowledge that their research has been conducted within the framework of the Trieste Institute for Theoretical Quantum Technologies (TQT).

\section*{Author contributions}
VV set the theoretical framework. AS performed the simulations and the experiments on IBM Quantum.
All authors discussed and contributed in writing the manuscript.

\appendix
\section{Derivation of LGI}\label{app:Derivation}
Let us derive the Leggett-Garg's inequalities in Eq.~(\ref{eq:LGI}) explicitly. Derivation of LGBI is analogous where Bell locality plays the role of NIM.\\
Let us start with the definition of a classical dichotomic variable $Q$. It must \new{take} two values $Q=\pm 1$, but it is not necessary associated to a dichotomic operator \cite{Lambert2016}. We use $Q(t_i)=Q_i$  to denote the measurement value of the observable at time $t_i$. Finally we label the probability of obtaining the result $Q_i$ at time $t_i$ as $P_i(Q_i)$. Therefore, the correlation function $C_{ij}$ can be defined as follows:
\begin{equation}
C_{ij}= \sum_{Q_{i},Q_{j}=\pm 1}Q_iQ_jP_{ij}(Q_i,Q_j),
\label{eq:1.1}
\end{equation}
where the subscripts of $P$ are used to explicitly remind the reader of the times at which the measurements were performed. The assumption of macrorealism per se guarantees that $P_{ij}$ can be obtained as the marginal probability of $P_{ij}(Q_i,Q_j,Q_k)$.
\begin{equation}\label{eq:Pij}
P_{ijk}=\sum_{Q_{k};k\neq i,j}P_{ij\new{k}}(Q_i,Q_j,Q_k)
\end{equation}
Without the assumption of non-invasive measurability, earlier measurements may affect the \new{following ones} and the probabilities do not necessarily come from a joint probability distribution. 
\new{Considering instead the NIM assumption to hold, measurements do not affect the state of the system or the subsequent system dynamics. Therefore the order of the measurements is not important; one can} drop the subscripts of $P_{ij\new{k}}$ and use the $P(Q_i,Q_j,Q_k)$ to calculate the three correlation functions: $C_{12},C_{23},C_{13}$.
Starting from the general expression 
\begin{equation}
C_{ij}=\sum_{Q_i,Q_j=\pm 1}Q_iQ_jP(Q_i,Q_j)= \langle Q_iQ_j \rangle,
\end{equation}
we obtain
\begin{equation}
\begin{aligned}
C_{12}=&P(+,+,+)+P(+,+,-)+P(-,-,+)+\\ &+P(-,-,-)-P(+,-,+)-P(+,-,-)+\\&-P(-,+,+)-P(-,+,-),\\
C_{13}=&P(+,+,+)+P(+,-,+)+P(-,+,-)+\\&+P(-,-,-)-P(+,+,-)-P(+,-,-)+\\&-P(-,+,+)-P(-,-,+),\\
C_{23}=&P(+,+,+)+P(-,+,+)+P(+,-,-)+\\&+P(-,-,-)-P(+,+,-)-P(-,+,+)+\\&-P(+,-,+)-P(-,-,+),
\end{aligned}
\label{eq:Cij}
\end{equation}
where we have used $P(\pm,\pm,\pm)=P(\pm 1,\pm 1,\pm 1)$.\\
Exploiting the completeness relation $\sum_{Q_{i},Q_{j},Q_{k}}P(Q_i,Q_j,Q_k)=1$, we obtain $K_3=C_{12}+C_{23}-C_{13}$:
\begin{equation}
K_3=1-4[P(+,-,+)+P(-,+,-)].
\end{equation}
The upper bound of $K_3$ is given by $P(+,-,+)=P(-,+,-)=0$ which is $K_3=1$; the lower bound, instead, is given by $P(+,-,+)+P(-,+,-)=1$, hence $K_3 \geq -3 $.
Besides the above inequality, that is
\begin{equation}
-3 \leq K_3 \leq 1
\end{equation}
other inequalities exist, that can be found in the literature.\\
Various symmetry properties may be used to derive further constrains on the correlations. First, we can redefine the dichotomic variable $Q \rightarrow -Q  $ at different times in $K_3$. This operation generates the following inequality:
\begin{equation}
-3\leq K'_3 \leq 1; \; \; \; K'_3 \equiv -C_{12} -C_{23} -C_{13}.
\end{equation} 
Finally, the last, different, third order inequality can be obtained from $K_3$, just changing a sign: 
\begin{equation}
-3\leq K_{3}^\mathrm{perm} \leq 1; \; \; \; K_{3}^\mathrm{perm} \equiv -C_{12} +C_{23} +C_{13}.
\end{equation}  
In principle, one can derive other functions starting from the function $K_3$ and building all the quantities obtained permuting all the time indices. 
In our example (three measurements in time, i.e. order $3$), the only three different cases are the inequalities proposed above.\\

\section{Error mitigation}\label{app:ErrorMitigation}

\begin{figure}
    \centering
    \includegraphics[width=\linewidth]{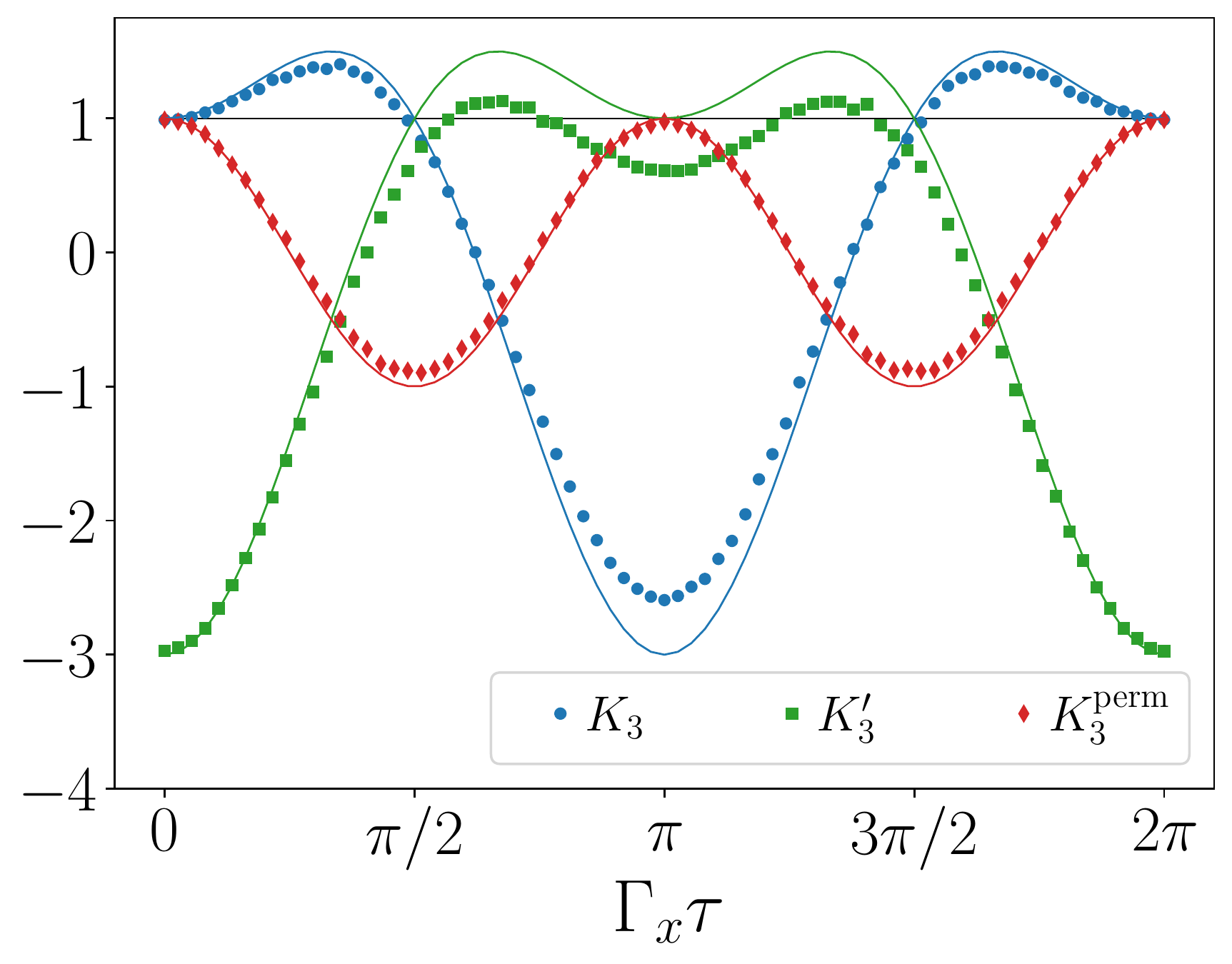}
    \caption{As in Fig.~\ref{fig:LGI_qubit} without error mitigation. LGI as a function of the time difference between two subsequent measurements.}
    \label{fig:noisyLGI}
\end{figure}
\begin{figure*}
    \begin{minipage}{\linewidth}
    \includegraphics[width=0.32\linewidth]{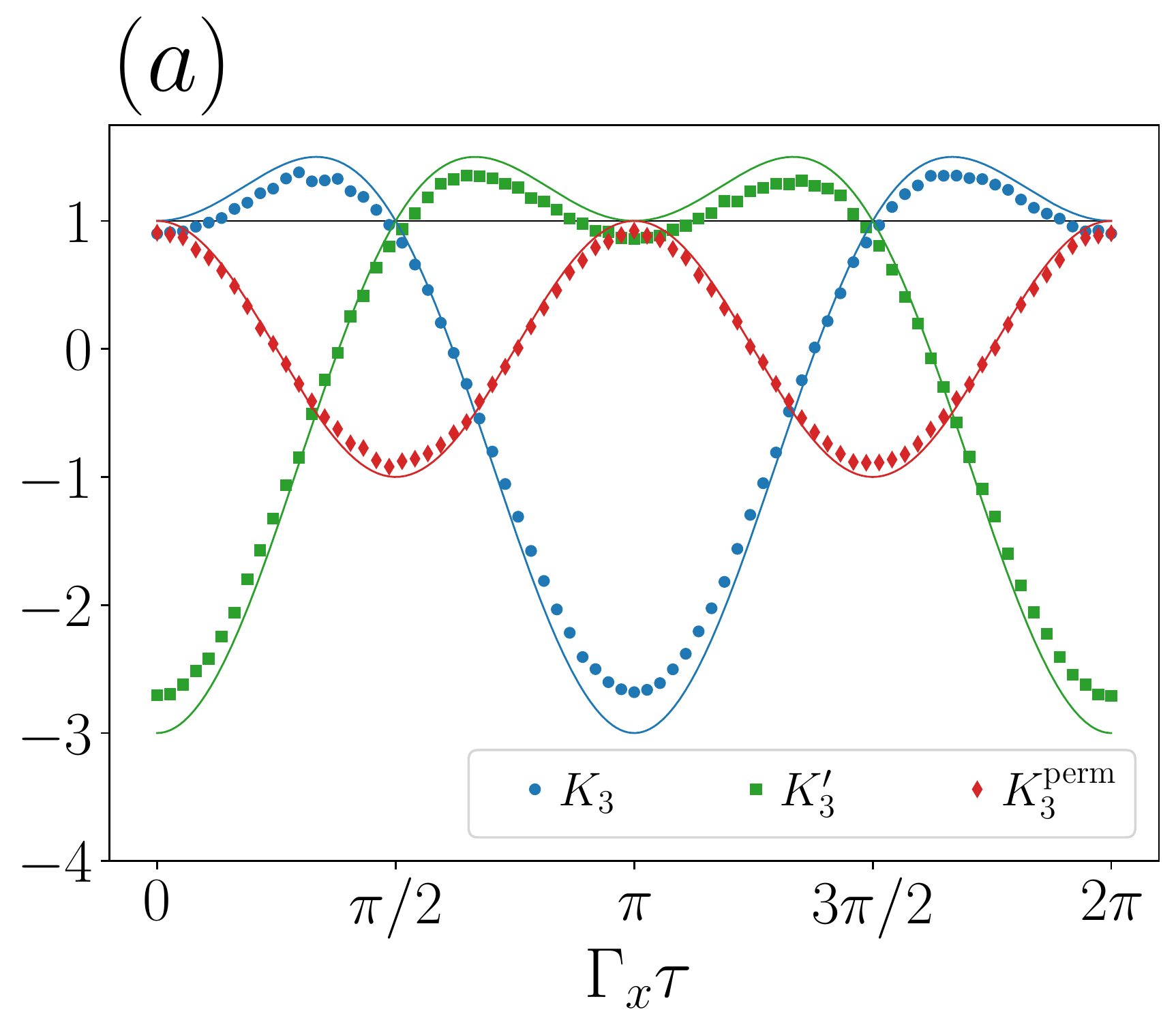}
    \includegraphics[width=0.32\linewidth]{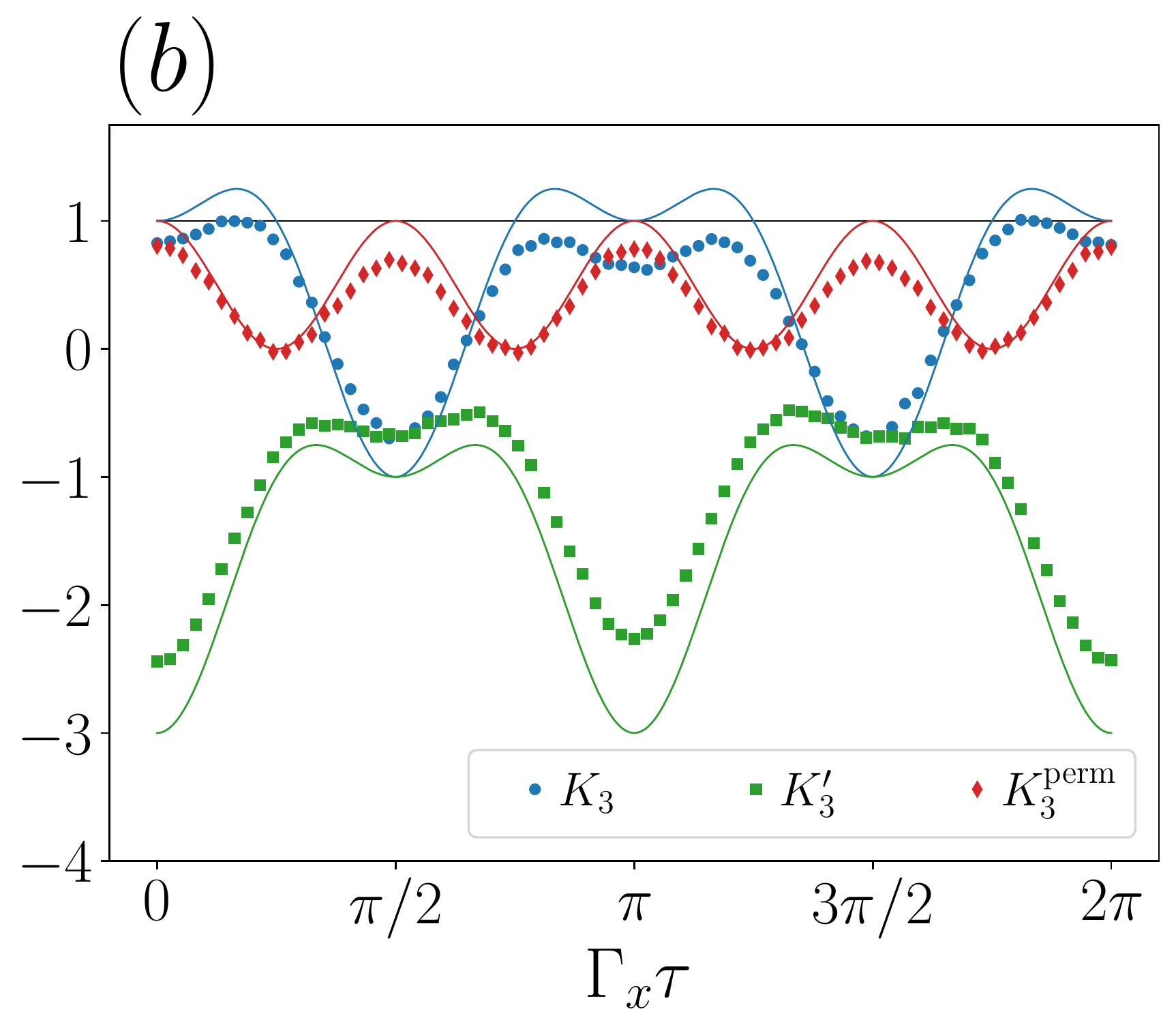}
    \includegraphics[width=0.32\linewidth]{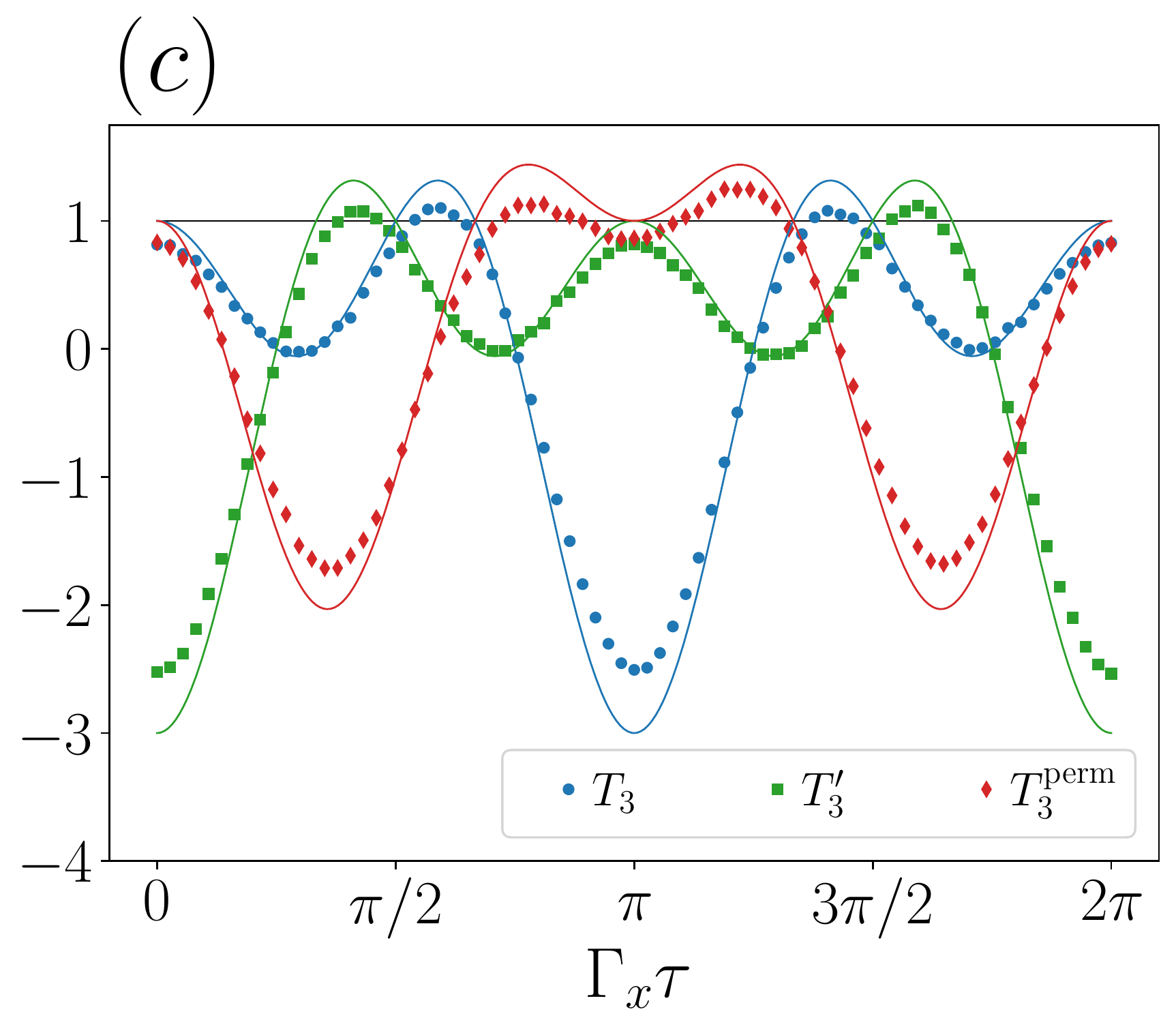}
    \end{minipage}
    \begin{minipage}{\linewidth}
    \includegraphics[width=\linewidth]{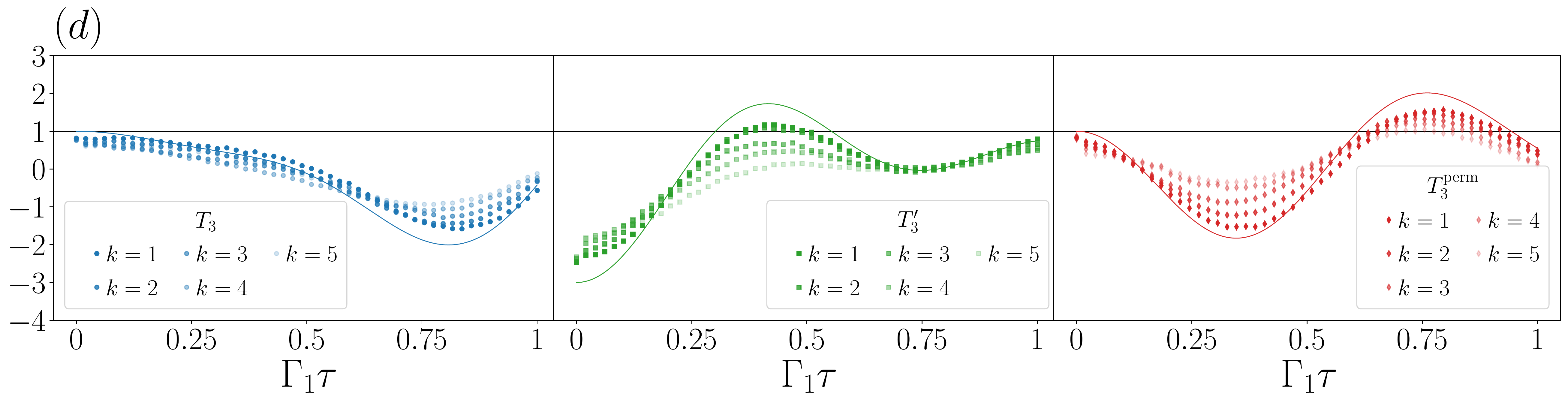}
    \end{minipage}
    \caption{\new{$(a)-(c)$ As in Fig.~\ref{fig:LGB_bellstate} without error mitigation, $(d)$ as in Fig.~\ref{fig:TFIC04} without error mitigation. $(a)-(c)$ LGBIs and LGIs on a two-qubits system. Markers are experimental results, solid lines are numerical predictions obtained with exact diagonalization. In panel $(a)$ LGIs calculated on a single qubit; in panel $(b)$ LGIs considering the total spin of the system; in panel $(c)$ LGBIs. $(d)$ LGBIs calculated for the Transverse field Ising chain. }}
    \label{fig:LGB_bellstate_noisy}
\end{figure*}

Quantum error mitigation refers to a series of techniques aimed at reducing (mitigating) the errors that occur in quantum computing algorithms due to hardware limitations.
These techniques try to reduce the impact of noise in quantum computations without, in general, completely removing it.
As the sources of noise during quantum computing algorithms are present through the whole computation and a complete account of these is a very complex task, we decided to deal only with the error occurring at readout. 
Namely, we tried to mitigate the readout noise of IBM Quantum, correcting the output of the experiment a posteriori.
A way to deal with it is already implemented in IBM Quantum and quite simple in its realization; here we give an account of the basics.
Let us assume to have an $n$-qubit system. We perform a measurement on all the qubits at a certain time and obtain an $n$-bit string output. This string could be one of $2^n$ possible realizations.  
Now let us assume that the same output is read again and a different outcome is collected. We might imagine that the noise perturbed the measurement outcome, slightly modifying the output string.
Easily enough one could prepare all possible $2^n$ states of the $n$-qubit string and perform a readout on all of them, in the hypothesis that the error on the preparation of the states is negligible.
\new{Repeating the readout \new{several} times on each of the $2^n$ possible states, one can construct a matrix $M$ which contains the outcome probabilities for each initial state, ideally without errors $M=\mathbb{I}$.}
In the framework we are working in, we may assume that the noisy measurement output $\sigma_{\textrm{noisy}}$ \new{is the product of $M$ times a noiseless output} $\sigma_{\textrm{noiseless}}$:
\begin{equation}
    \sigma_{\textrm{noisy}}=M\cdot\sigma_{\textrm{noiseless}}.
\end{equation}
The error mitigated results $\sigma_{\textrm{noiseless}}$ are obtained applying $M^{-1}$ on $\sigma_{\textrm{noisy}}$.

This approach allows to collect better results than the not-mitigated ones. We also acknowledge that its complexity scales exponentially with the system size making it necessary to deal with few qubits measurements.
In Fig.~\ref{fig:noisyLGI} we show an explicit example of experimental measurements without error mitigation. The results are a replica of the ones in Fig.~\ref{fig:LGI_qubit}. \new{When error mitigation is not performed, the performance of the hardware is evidently worse.}\\
\new{For completeness we also show the results without error mitigation for Fig. \ref{fig:LGB_bellstate} and Fig. \ref{fig:TFIC04} in Fig. \ref{fig:LGB_bellstate_noisy}.}

\begin{figure}
    \centering
    \includegraphics[width=\linewidth]{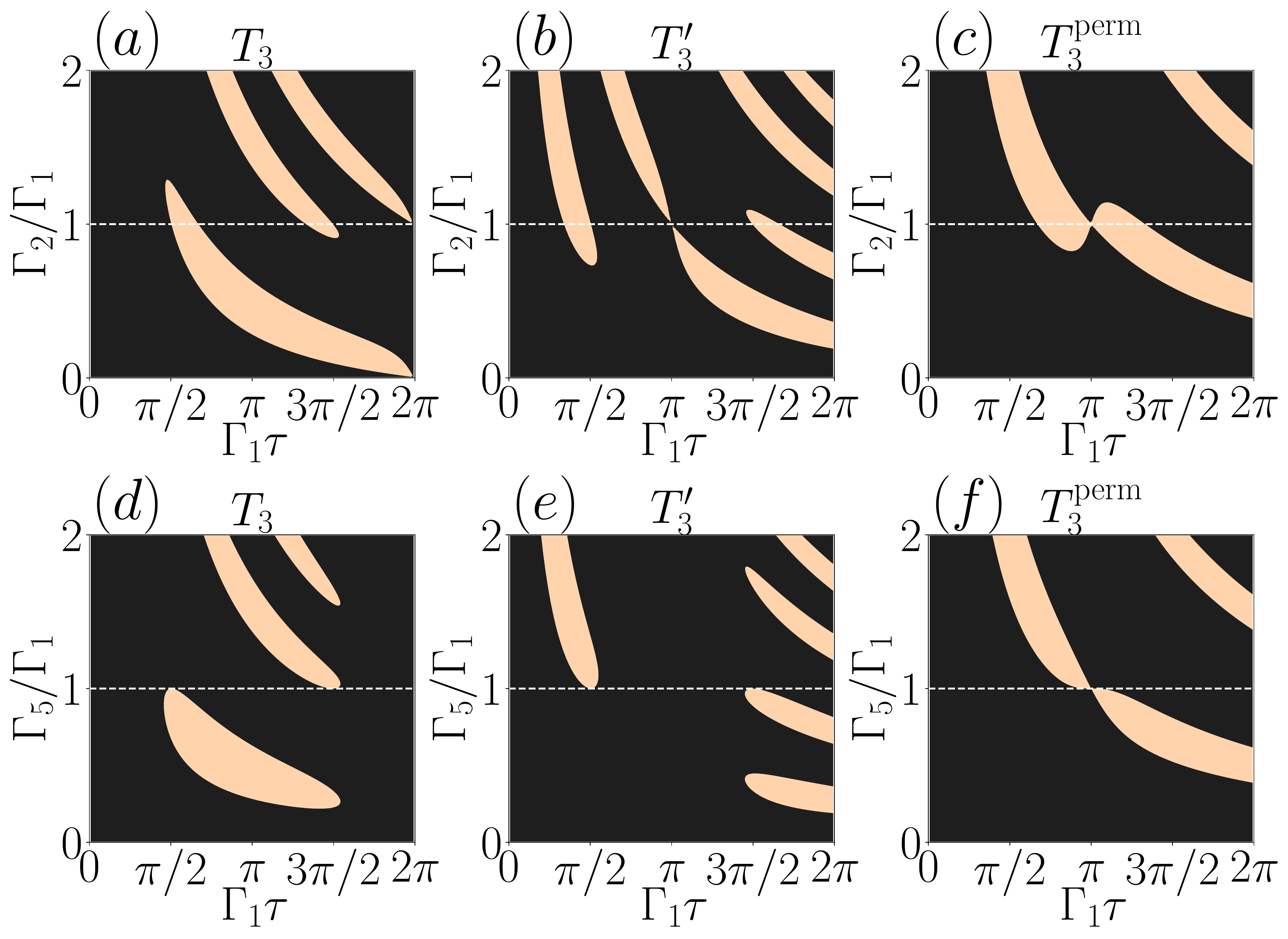}
    \caption{Contour plot of the LGBIs as a function of time and parameters $\Gamma_i$ for a system described by Eq.\ref{eq:appC}. The results are obtained via exact diagonalization according to Eq.\ref{eq:analyticalCij}. Yellow (black) areas correspond to regions where there is (not) a violation of the LGBIs. Upper panels: simulations for a system of 2 qubits. Lower panels: simulations for a system of 5 qubits.}
    \label{fig:noninteractingcplot}
\end{figure}

\new{
\section{Estimation of the parameters for the interacting evolution}\label{app:EPIE}
In this appendix we explain the choice of the parameters for the simulation of the many-body system, discussed in Sec. \ref{sec:multiqubit}.
We examine the dynamics of n-qubits, initially entangled in a GHZ state $(\ket{0}^{\otimes n} + \ket{1}^{\otimes n})/\sqrt{2}$, evolving independently according to the Hamiltonian
\begin{equation}\label{eq:appC}
H=\sum_{i=1}^N \frac{\Gamma_i}{2} \sigma^x_i,
\end{equation}
In particular in Fig.\ref{fig:noninteractingcplot} we show contour plots of  $T_3$, $T_3^{'}$ and $T_3^{\textrm{perm}}$ in two different cases: i) upper panels show the result of the LGBIs among qubit 1 and qubit 2 for a system composed of two qubits as a function of time and the ratio $\Gamma_2/\Gamma_1$; ii) lower panels show the same quantities among qubit 1 and qubit 5 evaluated for a system composed of five qubits as a function of $\Gamma_5/\Gamma_1$ with $\Gamma_{2,3,4}=\Gamma_1$.
The black regions correspond to values of the parameters where the LGBIs are not violated, yellow lobes correspond to violations of the inequalities.}

\new{We observe that, in the case of the two-qubits system, a violation of the inequalities occur  for any value of the ratio $\Gamma_2/\Gamma_1$ as a function of time.
On the other hand, in the case of the five-qubits system for $\Gamma_5/\Gamma_1=1$ no violation occurs in any of $T_3$, $T_3^{'}$ and $T_3^{\textrm{perm}}$.
For this reason, in Sec.\ref{sec:multiqubit}, we investigated the case where $\Gamma_5/\Gamma_1=2$ and $\Gamma_{2,3,4}=\Gamma_1$.}

\section{Hardware properties}
In this section we provide experimental details of the quantum device. In table \ref{tab:hardware_properties} we report the characterization data of the IBM Quantum processor at the time in which the simulations where performed. We refer the reader to the official IBM Quantum site for further details \cite{IBMQ_ref}. 
\begin{center}\begin{table}
\begin{tabular}{|c|c|}\hline Quantum hardware properties & Estimated value \\ \hline
         Average $\sqrt{X}, X, R_z(\theta)$ error & $0.0003$ \\ \hline
         Average $\mathrm{CNOT}$ gate error & $0.01$\\\hline
         Average readout error & $0.03$\\ \hline
         Average relaxation time qubits & $140\ \mu s $\\\hline
         Average decoherence time & $60\ \mu s$\\\hline
         Minimum decoherence/relaxation time & $24\ \mu s$\\
         \hline
         Gate length & $0.03\ \mu s$\\\hline 
\end{tabular}
\caption{Characterization data of `ibmq\_manila' quantum device at the time of the experiments.}\label{tab:hardware_properties}
\end{table}
\end{center}

\bibliography{biblio_resub.bib}

\begin{thebibliography}{49}%
\makeatletter
\providecommand \@ifxundefined [1]{%
 \@ifx{#1\undefined}
}%
\providecommand \@ifnum [1]{%
 \ifnum #1\expandafter \@firstoftwo
 \else \expandafter \@secondoftwo
 \fi
}%
\providecommand \@ifx [1]{%
 \ifx #1\expandafter \@firstoftwo
 \else \expandafter \@secondoftwo
 \fi
}%
\providecommand \natexlab [1]{#1}%
\providecommand \enquote  [1]{``#1''}%
\providecommand \bibnamefont  [1]{#1}%
\providecommand \bibfnamefont [1]{#1}%
\providecommand \citenamefont [1]{#1}%
\providecommand \href@noop [0]{\@secondoftwo}%
\providecommand \href [0]{\begingroup \@sanitize@url \@href}%
\providecommand \@href[1]{\@@startlink{#1}\@@href}%
\providecommand \@@href[1]{\endgroup#1\@@endlink}%
\providecommand \@sanitize@url [0]{\catcode `\\12\catcode `\$12\catcode
  `\&12\catcode `\#12\catcode `\^12\catcode `\_12\catcode `\%12\relax}%
\providecommand \@@startlink[1]{}%
\providecommand \@@endlink[0]{}%
\providecommand \url  [0]{\begingroup\@sanitize@url \@url }%
\providecommand \@url [1]{\endgroup\@href {#1}{\urlprefix }}%
\providecommand \urlprefix  [0]{URL }%
\providecommand \Eprint [0]{\href }%
\providecommand \doibase [0]{https://doi.org/}%
\providecommand \selectlanguage [0]{\@gobble}%
\providecommand \bibinfo  [0]{\@secondoftwo}%
\providecommand \bibfield  [0]{\@secondoftwo}%
\providecommand \translation [1]{[#1]}%
\providecommand \BibitemOpen [0]{}%
\providecommand \bibitemStop [0]{}%
\providecommand \bibitemNoStop [0]{.\EOS\space}%
\providecommand \EOS [0]{\spacefactor3000\relax}%
\providecommand \BibitemShut  [1]{\csname bibitem#1\endcsname}%
\let\auto@bib@innerbib\@empty
\bibitem [{\citenamefont {Preskill}(2018)}]{Preskill_2018}%
  \BibitemOpen
  \bibfield  {author} {\bibinfo {author} {\bibfnamefont {J.}~\bibnamefont
  {Preskill}},\ }\bibfield  {title} {\bibinfo {title} {Quantum computing in the
  nisq era and beyond},\ }\href {https://doi.org/10.22331/q-2018-08-06-79}
  {\bibfield  {journal} {\bibinfo  {journal} {Quantum}\ }\textbf {\bibinfo
  {volume} {2}},\ \bibinfo {pages} {79} (\bibinfo {year} {2018})}\BibitemShut
  {NoStop}%
\bibitem [{\citenamefont {Brydges}\ \emph {et~al.}(2019)\citenamefont
  {Brydges}, \citenamefont {Elben}, \citenamefont {Jurcevic}, \citenamefont
  {Vermersch}, \citenamefont {Maier}, \citenamefont {Lanyon}, \citenamefont
  {Zoller}, \citenamefont {Blatt},\ and\ \citenamefont
  {Roos}}]{brydges2019probing}%
  \BibitemOpen
  \bibfield  {author} {\bibinfo {author} {\bibfnamefont {T.}~\bibnamefont
  {Brydges}}, \bibinfo {author} {\bibfnamefont {A.}~\bibnamefont {Elben}},
  \bibinfo {author} {\bibfnamefont {P.}~\bibnamefont {Jurcevic}}, \bibinfo
  {author} {\bibfnamefont {B.}~\bibnamefont {Vermersch}}, \bibinfo {author}
  {\bibfnamefont {C.}~\bibnamefont {Maier}}, \bibinfo {author} {\bibfnamefont
  {B.~P.}\ \bibnamefont {Lanyon}}, \bibinfo {author} {\bibfnamefont
  {P.}~\bibnamefont {Zoller}}, \bibinfo {author} {\bibfnamefont
  {R.}~\bibnamefont {Blatt}},\ and\ \bibinfo {author} {\bibfnamefont {C.~F.}\
  \bibnamefont {Roos}},\ }\bibfield  {title} {\bibinfo {title} {Probing
  r{\'e}nyi entanglement entropy via randomized measurements},\ }\href
  {https://doi.org/10.1126/science.aau4963} {\bibfield  {journal} {\bibinfo
  {journal} {Science}\ }\textbf {\bibinfo {volume} {364}},\ \bibinfo {pages}
  {260} (\bibinfo {year} {2019})}\BibitemShut {NoStop}%
\bibitem [{\citenamefont {Vitale}\ \emph {et~al.}(2021)\citenamefont {Vitale},
  \citenamefont {Elben}, \citenamefont {Kueng}, \citenamefont {Neven},
  \citenamefont {Carrasco}, \citenamefont {Kraus}, \citenamefont {Zoller},
  \citenamefont {Calabrese}, \citenamefont {Vermersch},\ and\ \citenamefont
  {Dalmonte}}]{vitale2021symmetry}%
  \BibitemOpen
  \bibfield  {author} {\bibinfo {author} {\bibfnamefont {V.}~\bibnamefont
  {Vitale}}, \bibinfo {author} {\bibfnamefont {A.}~\bibnamefont {Elben}},
  \bibinfo {author} {\bibfnamefont {R.}~\bibnamefont {Kueng}}, \bibinfo
  {author} {\bibfnamefont {A.}~\bibnamefont {Neven}}, \bibinfo {author}
  {\bibfnamefont {J.}~\bibnamefont {Carrasco}}, \bibinfo {author}
  {\bibfnamefont {B.}~\bibnamefont {Kraus}}, \bibinfo {author} {\bibfnamefont
  {P.}~\bibnamefont {Zoller}}, \bibinfo {author} {\bibfnamefont
  {P.}~\bibnamefont {Calabrese}}, \bibinfo {author} {\bibfnamefont
  {B.}~\bibnamefont {Vermersch}},\ and\ \bibinfo {author} {\bibfnamefont
  {M.}~\bibnamefont {Dalmonte}},\ }\bibfield  {title} {\bibinfo {title}
  {Symmetry-resolved dynamical purification in synthetic quantum matter},\
  }\href {https://arxiv.org/abs/2101.07814} {\bibfield  {journal} {\bibinfo
  {journal} {arXiv preprint arXiv:2101.07814}\ } (\bibinfo {year}
  {2021})}\BibitemShut {NoStop}%
\bibitem [{\citenamefont {Elben}\ \emph {et~al.}(2020)\citenamefont {Elben},
  \citenamefont {Kueng}, \citenamefont {Huang}, \citenamefont {van Bijnen},
  \citenamefont {Kokail}, \citenamefont {Dalmonte}, \citenamefont {Calabrese},
  \citenamefont {Kraus}, \citenamefont {Preskill}, \citenamefont {Zoller} \emph
  {et~al.}}]{elben2020mixed}%
  \BibitemOpen
  \bibfield  {author} {\bibinfo {author} {\bibfnamefont {A.}~\bibnamefont
  {Elben}}, \bibinfo {author} {\bibfnamefont {R.}~\bibnamefont {Kueng}},
  \bibinfo {author} {\bibfnamefont {H.-Y.~R.}\ \bibnamefont {Huang}}, \bibinfo
  {author} {\bibfnamefont {R.}~\bibnamefont {van Bijnen}}, \bibinfo {author}
  {\bibfnamefont {C.}~\bibnamefont {Kokail}}, \bibinfo {author} {\bibfnamefont
  {M.}~\bibnamefont {Dalmonte}}, \bibinfo {author} {\bibfnamefont
  {P.}~\bibnamefont {Calabrese}}, \bibinfo {author} {\bibfnamefont
  {B.}~\bibnamefont {Kraus}}, \bibinfo {author} {\bibfnamefont
  {J.}~\bibnamefont {Preskill}}, \bibinfo {author} {\bibfnamefont
  {P.}~\bibnamefont {Zoller}}, \emph {et~al.},\ }\bibfield  {title} {\bibinfo
  {title} {Mixed-state entanglement from local randomized measurements},\
  }\href {https://journals.aps.org/prl/abstract/10.1103/PhysRevLett.125.200501}
  {\bibfield  {journal} {\bibinfo  {journal} {Physical Review Letters}\
  }\textbf {\bibinfo {volume} {125}},\ \bibinfo {pages} {200501} (\bibinfo
  {year} {2020})}\BibitemShut {NoStop}%
\bibitem [{\citenamefont {Neven}\ \emph {et~al.}(2021)\citenamefont {Neven},
  \citenamefont {Carrasco}, \citenamefont {Vitale}, \citenamefont {Kokail},
  \citenamefont {Elben}, \citenamefont {Dalmonte}, \citenamefont {Calabrese},
  \citenamefont {Zoller}, \citenamefont {Vermersch}, \citenamefont {Kueng}
  \emph {et~al.}}]{neven2021symmetry}%
  \BibitemOpen
  \bibfield  {author} {\bibinfo {author} {\bibfnamefont {A.}~\bibnamefont
  {Neven}}, \bibinfo {author} {\bibfnamefont {J.}~\bibnamefont {Carrasco}},
  \bibinfo {author} {\bibfnamefont {V.}~\bibnamefont {Vitale}}, \bibinfo
  {author} {\bibfnamefont {C.}~\bibnamefont {Kokail}}, \bibinfo {author}
  {\bibfnamefont {A.}~\bibnamefont {Elben}}, \bibinfo {author} {\bibfnamefont
  {M.}~\bibnamefont {Dalmonte}}, \bibinfo {author} {\bibfnamefont
  {P.}~\bibnamefont {Calabrese}}, \bibinfo {author} {\bibfnamefont
  {P.}~\bibnamefont {Zoller}}, \bibinfo {author} {\bibfnamefont
  {B.}~\bibnamefont {Vermersch}}, \bibinfo {author} {\bibfnamefont
  {R.}~\bibnamefont {Kueng}}, \emph {et~al.},\ }\bibfield  {title} {\bibinfo
  {title} {Symmetry-resolved entanglement detection using partial transpose
  moments},\ }\href {https://arxiv.org/abs/2103.07443} {\bibfield  {journal}
  {\bibinfo  {journal} {arXiv preprint arXiv:2103.07443}\ } (\bibinfo {year}
  {2021})}\BibitemShut {NoStop}%
\bibitem [{\citenamefont {Yu}\ \emph {et~al.}(2021)\citenamefont {Yu},
  \citenamefont {Imai},\ and\ \citenamefont {G{\"u}hne}}]{yu2021optimal}%
  \BibitemOpen
  \bibfield  {author} {\bibinfo {author} {\bibfnamefont {X.-D.}\ \bibnamefont
  {Yu}}, \bibinfo {author} {\bibfnamefont {S.}~\bibnamefont {Imai}},\ and\
  \bibinfo {author} {\bibfnamefont {O.}~\bibnamefont {G{\"u}hne}},\ }\bibfield
  {title} {\bibinfo {title} {Optimal entanglement certification from moments of
  the partial transpose},\ }\href {https://arxiv.org/abs/2103.06897} {\bibfield
   {journal} {\bibinfo  {journal} {arXiv preprint arXiv:2103.06897}\ }
  (\bibinfo {year} {2021})}\BibitemShut {NoStop}%
\bibitem [{\citenamefont {Leggett}\ and\ \citenamefont {Garg}(1985)}]{LGI}%
  \BibitemOpen
  \bibfield  {author} {\bibinfo {author} {\bibfnamefont {A.~J.}\ \bibnamefont
  {Leggett}}\ and\ \bibinfo {author} {\bibfnamefont {A.}~\bibnamefont {Garg}},\
  }\bibfield  {title} {\bibinfo {title} {Quantum mechanics versus macroscopic
  realism: Is the flux there when nobody looks?},\ }\href
  {https://doi.org/10.1103/PhysRevLett.54.857} {\bibfield  {journal} {\bibinfo
  {journal} {Phys. Rev. Lett.}\ }\textbf {\bibinfo {volume} {54}},\ \bibinfo
  {pages} {857} (\bibinfo {year} {1985})}\BibitemShut {NoStop}%
\bibitem [{\citenamefont {Leggett}(2002)}]{Leggett_2002}%
  \BibitemOpen
  \bibfield  {author} {\bibinfo {author} {\bibfnamefont {A.~J.}\ \bibnamefont
  {Leggett}},\ }\href {https://doi.org/10.1088/0953-8984/14/15/201} {\ \textbf
  {\bibinfo {volume} {14}},\ \bibinfo {pages} {R415} (\bibinfo {year}
  {2002})}\BibitemShut {NoStop}%
\bibitem [{\citenamefont {IBM-Quantum}(2021)}]{IBMQ_ref}%
  \BibitemOpen
  \bibfield  {author} {\bibinfo {author} {\bibnamefont {IBM-Quantum}},\ }\href
  {https://doi.org/https://quantum-computing.ibm.com/} {\bibinfo {title}
  {https://quantum-computing.ibm.com/}} (\bibinfo {year} {2021})\BibitemShut
  {NoStop}%
\bibitem [{\citenamefont {Palacios-Laloy}\ \emph {et~al.}(2010)\citenamefont
  {Palacios-Laloy}, \citenamefont {Mallet}, \citenamefont {Nguyen},
  \citenamefont {Bertet}, \citenamefont {Vion}, \citenamefont {Esteve},\ and\
  \citenamefont {Korotkov}}]{palacios2010experimental}%
  \BibitemOpen
  \bibfield  {author} {\bibinfo {author} {\bibfnamefont {A.}~\bibnamefont
  {Palacios-Laloy}}, \bibinfo {author} {\bibfnamefont {F.}~\bibnamefont
  {Mallet}}, \bibinfo {author} {\bibfnamefont {F.}~\bibnamefont {Nguyen}},
  \bibinfo {author} {\bibfnamefont {P.}~\bibnamefont {Bertet}}, \bibinfo
  {author} {\bibfnamefont {D.}~\bibnamefont {Vion}}, \bibinfo {author}
  {\bibfnamefont {D.}~\bibnamefont {Esteve}},\ and\ \bibinfo {author}
  {\bibfnamefont {A.~N.}\ \bibnamefont {Korotkov}},\ }\bibfield  {title}
  {\bibinfo {title} {Experimental violation of a bell’s inequality in time
  with weak measurement},\ }\href {https://www.nature.com/articles/nphys1641}
  {\bibfield  {journal} {\bibinfo  {journal} {Nature Physics}\ }\textbf
  {\bibinfo {volume} {6}},\ \bibinfo {pages} {442} (\bibinfo {year}
  {2010})}\BibitemShut {NoStop}%
\bibitem [{\citenamefont {Knee}\ \emph {et~al.}(2012)\citenamefont {Knee},
  \citenamefont {Simmons}, \citenamefont {Gauger}, \citenamefont {Morton},
  \citenamefont {Riemann}, \citenamefont {Abrosimov}, \citenamefont {Becker},
  \citenamefont {Pohl}, \citenamefont {Itoh}, \citenamefont {Thewalt} \emph
  {et~al.}}]{knee2012violation}%
  \BibitemOpen
  \bibfield  {author} {\bibinfo {author} {\bibfnamefont {G.~C.}\ \bibnamefont
  {Knee}}, \bibinfo {author} {\bibfnamefont {S.}~\bibnamefont {Simmons}},
  \bibinfo {author} {\bibfnamefont {E.~M.}\ \bibnamefont {Gauger}}, \bibinfo
  {author} {\bibfnamefont {J.~J.}\ \bibnamefont {Morton}}, \bibinfo {author}
  {\bibfnamefont {H.}~\bibnamefont {Riemann}}, \bibinfo {author} {\bibfnamefont
  {N.~V.}\ \bibnamefont {Abrosimov}}, \bibinfo {author} {\bibfnamefont
  {P.}~\bibnamefont {Becker}}, \bibinfo {author} {\bibfnamefont {H.-J.}\
  \bibnamefont {Pohl}}, \bibinfo {author} {\bibfnamefont {K.~M.}\ \bibnamefont
  {Itoh}}, \bibinfo {author} {\bibfnamefont {M.~L.}\ \bibnamefont {Thewalt}},
  \emph {et~al.},\ }\bibfield  {title} {\bibinfo {title} {Violation of a
  leggett--garg inequality with ideal non-invasive measurements},\ }\href
  {https://www.nature.com/articles/ncomms1614} {\bibfield  {journal} {\bibinfo
  {journal} {Nature communications}\ }\textbf {\bibinfo {volume} {3}},\
  \bibinfo {pages} {1} (\bibinfo {year} {2012})}\BibitemShut {NoStop}%
\bibitem [{\citenamefont {Xu}\ \emph {et~al.}(2011)\citenamefont {Xu},
  \citenamefont {Li}, \citenamefont {Zou},\ and\ \citenamefont
  {Guo}}]{xu2011experimental}%
  \BibitemOpen
  \bibfield  {author} {\bibinfo {author} {\bibfnamefont {J.-S.}\ \bibnamefont
  {Xu}}, \bibinfo {author} {\bibfnamefont {C.-F.}\ \bibnamefont {Li}}, \bibinfo
  {author} {\bibfnamefont {X.-B.}\ \bibnamefont {Zou}},\ and\ \bibinfo {author}
  {\bibfnamefont {G.-C.}\ \bibnamefont {Guo}},\ }\bibfield  {title} {\bibinfo
  {title} {Experimental violation of the leggett-garg inequality under
  decoherence},\ }\href {https://www.nature.com/articles/srep00101} {\bibfield
  {journal} {\bibinfo  {journal} {Scientific reports}\ }\textbf {\bibinfo
  {volume} {1}},\ \bibinfo {pages} {1} (\bibinfo {year} {2011})}\BibitemShut
  {NoStop}%
\bibitem [{\citenamefont {Goggin}\ \emph {et~al.}(2011)\citenamefont {Goggin},
  \citenamefont {Almeida}, \citenamefont {Barbieri}, \citenamefont {Lanyon},
  \citenamefont {O’brien}, \citenamefont {White},\ and\ \citenamefont
  {Pryde}}]{goggin2011violation}%
  \BibitemOpen
  \bibfield  {author} {\bibinfo {author} {\bibfnamefont {M.}~\bibnamefont
  {Goggin}}, \bibinfo {author} {\bibfnamefont {M.}~\bibnamefont {Almeida}},
  \bibinfo {author} {\bibfnamefont {M.}~\bibnamefont {Barbieri}}, \bibinfo
  {author} {\bibfnamefont {B.}~\bibnamefont {Lanyon}}, \bibinfo {author}
  {\bibfnamefont {J.}~\bibnamefont {O’brien}}, \bibinfo {author}
  {\bibfnamefont {A.}~\bibnamefont {White}},\ and\ \bibinfo {author}
  {\bibfnamefont {G.}~\bibnamefont {Pryde}},\ }\bibfield  {title} {\bibinfo
  {title} {Violation of the leggett--garg inequality with weak measurements of
  photons},\ }\href {https://www.pnas.org/content/108/4/1256} {\bibfield
  {journal} {\bibinfo  {journal} {Proceedings of the National Academy of
  Sciences}\ }\textbf {\bibinfo {volume} {108}},\ \bibinfo {pages} {1256}
  (\bibinfo {year} {2011})}\BibitemShut {NoStop}%
\bibitem [{\citenamefont {Groen}\ \emph {et~al.}(2013)\citenamefont {Groen},
  \citenamefont {Rist\`e}, \citenamefont {Tornberg}, \citenamefont {Cramer},
  \citenamefont {de~Groot}, \citenamefont {Picot}, \citenamefont {Johansson},\
  and\ \citenamefont {DiCarlo}}]{Groen2013}%
  \BibitemOpen
  \bibfield  {author} {\bibinfo {author} {\bibfnamefont {J.~P.}\ \bibnamefont
  {Groen}}, \bibinfo {author} {\bibfnamefont {D.}~\bibnamefont {Rist\`e}},
  \bibinfo {author} {\bibfnamefont {L.}~\bibnamefont {Tornberg}}, \bibinfo
  {author} {\bibfnamefont {J.}~\bibnamefont {Cramer}}, \bibinfo {author}
  {\bibfnamefont {P.~C.}\ \bibnamefont {de~Groot}}, \bibinfo {author}
  {\bibfnamefont {T.}~\bibnamefont {Picot}}, \bibinfo {author} {\bibfnamefont
  {G.}~\bibnamefont {Johansson}},\ and\ \bibinfo {author} {\bibfnamefont
  {L.}~\bibnamefont {DiCarlo}},\ }\bibfield  {title} {\bibinfo {title}
  {Partial-measurement backaction and nonclassical weak values in a
  superconducting circuit},\ }\href
  {https://doi.org/10.1103/PhysRevLett.111.090506} {\bibfield  {journal}
  {\bibinfo  {journal} {Phys. Rev. Lett.}\ }\textbf {\bibinfo {volume} {111}},\
  \bibinfo {pages} {090506} (\bibinfo {year} {2013})}\BibitemShut {NoStop}%
\bibitem [{\citenamefont {Emary}\ \emph {et~al.}(2014)\citenamefont {Emary},
  \citenamefont {Cotter},\ and\ \citenamefont {Arndt}}]{Emary2014}%
  \BibitemOpen
  \bibfield  {author} {\bibinfo {author} {\bibfnamefont {C.}~\bibnamefont
  {Emary}}, \bibinfo {author} {\bibfnamefont {J.~P.}\ \bibnamefont {Cotter}},\
  and\ \bibinfo {author} {\bibfnamefont {M.}~\bibnamefont {Arndt}},\ }\bibfield
   {title} {\bibinfo {title} {Testing macroscopic realism through high-mass
  interferometry},\ }\href {https://doi.org/10.1103/PhysRevA.90.042114}
  {\bibfield  {journal} {\bibinfo  {journal} {Phys. Rev. A}\ }\textbf {\bibinfo
  {volume} {90}},\ \bibinfo {pages} {042114} (\bibinfo {year}
  {2014})}\BibitemShut {NoStop}%
\bibitem [{\citenamefont {Budroni}\ \emph {et~al.}(2015)\citenamefont
  {Budroni}, \citenamefont {Vitagliano}, \citenamefont {Colangelo},
  \citenamefont {Sewell}, \citenamefont {G\"uhne}, \citenamefont {T\'oth},\
  and\ \citenamefont {Mitchell}}]{Budroni2015}%
  \BibitemOpen
  \bibfield  {author} {\bibinfo {author} {\bibfnamefont {C.}~\bibnamefont
  {Budroni}}, \bibinfo {author} {\bibfnamefont {G.}~\bibnamefont {Vitagliano}},
  \bibinfo {author} {\bibfnamefont {G.}~\bibnamefont {Colangelo}}, \bibinfo
  {author} {\bibfnamefont {R.~J.}\ \bibnamefont {Sewell}}, \bibinfo {author}
  {\bibfnamefont {O.}~\bibnamefont {G\"uhne}}, \bibinfo {author} {\bibfnamefont
  {G.}~\bibnamefont {T\'oth}},\ and\ \bibinfo {author} {\bibfnamefont {M.~W.}\
  \bibnamefont {Mitchell}},\ }\bibfield  {title} {\bibinfo {title} {Quantum
  nondemolition measurement enables macroscopic leggett-garg tests},\ }\href
  {https://doi.org/10.1103/PhysRevLett.115.200403} {\bibfield  {journal}
  {\bibinfo  {journal} {Phys. Rev. Lett.}\ }\textbf {\bibinfo {volume} {115}},\
  \bibinfo {pages} {200403} (\bibinfo {year} {2015})}\BibitemShut {NoStop}%
\bibitem [{\citenamefont {Katiyar}\ \emph {et~al.}(2017)\citenamefont
  {Katiyar}, \citenamefont {Brodutch}, \citenamefont {Lu},\ and\ \citenamefont
  {Laflamme}}]{katiyar2017experimental}%
  \BibitemOpen
  \bibfield  {author} {\bibinfo {author} {\bibfnamefont {H.}~\bibnamefont
  {Katiyar}}, \bibinfo {author} {\bibfnamefont {A.}~\bibnamefont {Brodutch}},
  \bibinfo {author} {\bibfnamefont {D.}~\bibnamefont {Lu}},\ and\ \bibinfo
  {author} {\bibfnamefont {R.}~\bibnamefont {Laflamme}},\ }\bibfield  {title}
  {\bibinfo {title} {Experimental violation of the leggett--garg inequality in
  a three-level system},\ }\href
  {https://iopscience.iop.org/article/10.1088/1367-2630/aa5c51} {\bibfield
  {journal} {\bibinfo  {journal} {New Journal of Physics}\ }\textbf {\bibinfo
  {volume} {19}},\ \bibinfo {pages} {023033} (\bibinfo {year}
  {2017})}\BibitemShut {NoStop}%
\bibitem [{\citenamefont {Wang}\ \emph {et~al.}(2017)\citenamefont {Wang},
  \citenamefont {Emary}, \citenamefont {Zhan}, \citenamefont {Bian},
  \citenamefont {Li},\ and\ \citenamefont {Xue}}]{wang2017enhanced}%
  \BibitemOpen
  \bibfield  {author} {\bibinfo {author} {\bibfnamefont {K.}~\bibnamefont
  {Wang}}, \bibinfo {author} {\bibfnamefont {C.}~\bibnamefont {Emary}},
  \bibinfo {author} {\bibfnamefont {X.}~\bibnamefont {Zhan}}, \bibinfo {author}
  {\bibfnamefont {Z.}~\bibnamefont {Bian}}, \bibinfo {author} {\bibfnamefont
  {J.}~\bibnamefont {Li}},\ and\ \bibinfo {author} {\bibfnamefont
  {P.}~\bibnamefont {Xue}},\ }\bibfield  {title} {\bibinfo {title} {Enhanced
  violations of leggett-garg inequalities in an experimental three-level
  system},\ }\href
  {https://www.osapublishing.org/oe/fulltext.cfm?uri=oe-25-25-31462&id=379231}
  {\bibfield  {journal} {\bibinfo  {journal} {Optics express}\ }\textbf
  {\bibinfo {volume} {25}},\ \bibinfo {pages} {31462} (\bibinfo {year}
  {2017})}\BibitemShut {NoStop}%
\bibitem [{\citenamefont {Bose}\ \emph {et~al.}(2018)\citenamefont {Bose},
  \citenamefont {Home},\ and\ \citenamefont {Mal}}]{Bose2018}%
  \BibitemOpen
  \bibfield  {author} {\bibinfo {author} {\bibfnamefont {S.}~\bibnamefont
  {Bose}}, \bibinfo {author} {\bibfnamefont {D.}~\bibnamefont {Home}},\ and\
  \bibinfo {author} {\bibfnamefont {S.}~\bibnamefont {Mal}},\ }\bibfield
  {title} {\bibinfo {title} {Nonclassicality of the harmonic-oscillator
  coherent state persisting up to the macroscopic domain},\ }\href
  {https://doi.org/10.1103/PhysRevLett.120.210402} {\bibfield  {journal}
  {\bibinfo  {journal} {Phys. Rev. Lett.}\ }\textbf {\bibinfo {volume} {120}},\
  \bibinfo {pages} {210402} (\bibinfo {year} {2018})}\BibitemShut {NoStop}%
\bibitem [{\citenamefont {Friedenberger}\ and\ \citenamefont
  {Lutz}(2017)}]{Friedenberger2017}%
  \BibitemOpen
  \bibfield  {author} {\bibinfo {author} {\bibfnamefont {A.}~\bibnamefont
  {Friedenberger}}\ and\ \bibinfo {author} {\bibfnamefont {E.}~\bibnamefont
  {Lutz}},\ }\bibfield  {title} {\bibinfo {title} {Assessing the quantumness of
  a damped two-level system},\ }\href
  {https://doi.org/10.1103/PhysRevA.95.022101} {\bibfield  {journal} {\bibinfo
  {journal} {Phys. Rev. A}\ }\textbf {\bibinfo {volume} {95}},\ \bibinfo
  {pages} {022101} (\bibinfo {year} {2017})}\BibitemShut {NoStop}%
\bibitem [{\citenamefont {Vitale}\ \emph {et~al.}(2019)\citenamefont {Vitale},
  \citenamefont {De~Filippis}, \citenamefont {De~Candia}, \citenamefont
  {Tagliacozzo}, \citenamefont {Cataudella},\ and\ \citenamefont
  {Lucignano}}]{vitale2019assessing}%
  \BibitemOpen
  \bibfield  {author} {\bibinfo {author} {\bibfnamefont {V.}~\bibnamefont
  {Vitale}}, \bibinfo {author} {\bibfnamefont {G.}~\bibnamefont {De~Filippis}},
  \bibinfo {author} {\bibfnamefont {A.}~\bibnamefont {De~Candia}}, \bibinfo
  {author} {\bibfnamefont {A.}~\bibnamefont {Tagliacozzo}}, \bibinfo {author}
  {\bibfnamefont {V.}~\bibnamefont {Cataudella}},\ and\ \bibinfo {author}
  {\bibfnamefont {P.}~\bibnamefont {Lucignano}},\ }\bibfield  {title} {\bibinfo
  {title} {Assessing the quantumness of the annealing dynamics via leggett
  garg’s inequalities: a weak measurement approach},\ }\href
  {https://doi.org/10.1038/s41598-019-50081-8} {\bibfield  {journal} {\bibinfo
  {journal} {Scientific reports}\ }\textbf {\bibinfo {volume} {9}},\ \bibinfo
  {pages} {1} (\bibinfo {year} {2019})}\BibitemShut {NoStop}%
\bibitem [{\citenamefont {Huffman}\ and\ \citenamefont
  {Mizel}(2017)}]{Huffman2017}%
  \BibitemOpen
  \bibfield  {author} {\bibinfo {author} {\bibfnamefont {E.}~\bibnamefont
  {Huffman}}\ and\ \bibinfo {author} {\bibfnamefont {A.}~\bibnamefont
  {Mizel}},\ }\bibfield  {title} {\bibinfo {title} {Violation of noninvasive
  macrorealism by a superconducting qubit: Implementation of a leggett-garg
  test that addresses the clumsiness loophole},\ }\href
  {https://doi.org/10.1103/PhysRevA.95.032131} {\bibfield  {journal} {\bibinfo
  {journal} {Phys. Rev. A}\ }\textbf {\bibinfo {volume} {95}},\ \bibinfo
  {pages} {032131} (\bibinfo {year} {2017})}\BibitemShut {NoStop}%
\bibitem [{\citenamefont {Ku}\ \emph {et~al.}(2020)\citenamefont {Ku},
  \citenamefont {Lambert}, \citenamefont {Chan}, \citenamefont {Emary},
  \citenamefont {Chen},\ and\ \citenamefont {Nori}}]{ku2020experimental}%
  \BibitemOpen
  \bibfield  {author} {\bibinfo {author} {\bibfnamefont {H.-Y.}\ \bibnamefont
  {Ku}}, \bibinfo {author} {\bibfnamefont {N.}~\bibnamefont {Lambert}},
  \bibinfo {author} {\bibfnamefont {F.-J.}\ \bibnamefont {Chan}}, \bibinfo
  {author} {\bibfnamefont {C.}~\bibnamefont {Emary}}, \bibinfo {author}
  {\bibfnamefont {Y.-N.}\ \bibnamefont {Chen}},\ and\ \bibinfo {author}
  {\bibfnamefont {F.}~\bibnamefont {Nori}},\ }\bibfield  {title} {\bibinfo
  {title} {Experimental test of non-macrorealistic cat states in the cloud},\
  }\href {https://www.nature.com/articles/s41534-020-00321-x} {\bibfield
  {journal} {\bibinfo  {journal} {npj Quantum Information}\ }\textbf {\bibinfo
  {volume} {6}},\ \bibinfo {pages} {1} (\bibinfo {year} {2020})}\BibitemShut
  {NoStop}%
\bibitem [{\citenamefont {Solfanelli}\ \emph {et~al.}(2021)\citenamefont
  {Solfanelli}, \citenamefont {Santini},\ and\ \citenamefont
  {Campisi}}]{Solfanelli21Arxiv}%
  \BibitemOpen
  \bibfield  {author} {\bibinfo {author} {\bibfnamefont {A.}~\bibnamefont
  {Solfanelli}}, \bibinfo {author} {\bibfnamefont {A.}~\bibnamefont
  {Santini}},\ and\ \bibinfo {author} {\bibfnamefont {M.}~\bibnamefont
  {Campisi}},\ }\bibfield  {title} {\bibinfo {title} {Experimental verification
  of fluctuation relations with a quantum computer},\ }\href
  {https://doi.org/10.1103/PRXQuantum.2.030353} {\bibfield  {journal} {\bibinfo
   {journal} {PRX Quantum}\ }\textbf {\bibinfo {volume} {2}},\ \bibinfo {pages}
  {030353} (\bibinfo {year} {2021})}\BibitemShut {NoStop}%
\bibitem [{\citenamefont {Tacchino}\ \emph {et~al.}(2020)\citenamefont
  {Tacchino}, \citenamefont {Chiesa}, \citenamefont {Carretta},\ and\
  \citenamefont {Gerace}}]{Tacchino20AQT}%
  \BibitemOpen
  \bibfield  {author} {\bibinfo {author} {\bibfnamefont {F.}~\bibnamefont
  {Tacchino}}, \bibinfo {author} {\bibfnamefont {A.}~\bibnamefont {Chiesa}},
  \bibinfo {author} {\bibfnamefont {S.}~\bibnamefont {Carretta}},\ and\
  \bibinfo {author} {\bibfnamefont {D.}~\bibnamefont {Gerace}},\ }\bibfield
  {title} {\bibinfo {title} {Quantum computers as universal quantum simulators:
  State-of-the-art and perspectives},\ }\href
  {https://doi.org/https://doi.org/10.1002/qute.201900052} {\bibfield
  {journal} {\bibinfo  {journal} {Advanced Quantum Technologies}\ }\textbf
  {\bibinfo {volume} {3}},\ \bibinfo {pages} {1900052} (\bibinfo {year}
  {2020})}\BibitemShut {NoStop}%
\bibitem [{\citenamefont {Devitt}(2016)}]{Devitt2016}%
  \BibitemOpen
  \bibfield  {author} {\bibinfo {author} {\bibfnamefont {S.~J.}\ \bibnamefont
  {Devitt}},\ }\bibfield  {title} {\bibinfo {title} {Performing quantum
  computing experiments in the cloud},\ }\href
  {https://doi.org/10.1103/PhysRevA.94.032329} {\bibfield  {journal} {\bibinfo
  {journal} {Phys. Rev. A}\ }\textbf {\bibinfo {volume} {94}},\ \bibinfo
  {pages} {032329} (\bibinfo {year} {2016})}\BibitemShut {NoStop}%
\bibitem [{\citenamefont {Bravyi}\ \emph {et~al.}(2021)\citenamefont {Bravyi},
  \citenamefont {Sheldon}, \citenamefont {Kandala}, \citenamefont {Mckay},\
  and\ \citenamefont {Gambetta}}]{Bravyi2021}%
  \BibitemOpen
  \bibfield  {author} {\bibinfo {author} {\bibfnamefont {S.}~\bibnamefont
  {Bravyi}}, \bibinfo {author} {\bibfnamefont {S.}~\bibnamefont {Sheldon}},
  \bibinfo {author} {\bibfnamefont {A.}~\bibnamefont {Kandala}}, \bibinfo
  {author} {\bibfnamefont {D.~C.}\ \bibnamefont {Mckay}},\ and\ \bibinfo
  {author} {\bibfnamefont {J.~M.}\ \bibnamefont {Gambetta}},\ }\bibfield
  {title} {\bibinfo {title} {Mitigating measurement errors in multiqubit
  experiments},\ }\href {https://doi.org/10.1103/PhysRevA.103.042605}
  {\bibfield  {journal} {\bibinfo  {journal} {Phys. Rev. A}\ }\textbf {\bibinfo
  {volume} {103}},\ \bibinfo {pages} {042605} (\bibinfo {year}
  {2021})}\BibitemShut {NoStop}%
\bibitem [{\citenamefont {Jurcevic}\ \emph {et~al.}(2021)\citenamefont
  {Jurcevic}, \citenamefont {Javadi-Abhari}, \citenamefont {Bishop},
  \citenamefont {Lauer}, \citenamefont {Bogorin}, \citenamefont {Brink},
  \citenamefont {Capelluto}, \citenamefont {Günlük}, \citenamefont {Itoko},
  \citenamefont {Kanazawa}, \citenamefont {Kandala}, \citenamefont {Keefe},
  \citenamefont {Krsulich}, \citenamefont {Landers}, \citenamefont
  {Lewandowski}, \citenamefont {McClure}, \citenamefont {Nannicini},
  \citenamefont {Narasgond}, \citenamefont {Nayfeh}, \citenamefont {Pritchett},
  \citenamefont {Rothwell}, \citenamefont {Srinivasan}, \citenamefont
  {Sundaresan}, \citenamefont {Wang}, \citenamefont {Wei}, \citenamefont
  {Wood}, \citenamefont {Yau}, \citenamefont {Zhang}, \citenamefont {Dial},
  \citenamefont {Chow},\ and\ \citenamefont {Gambetta}}]{Jurcevic_2021}%
  \BibitemOpen
  \bibfield  {author} {\bibinfo {author} {\bibfnamefont {P.}~\bibnamefont
  {Jurcevic}}, \bibinfo {author} {\bibfnamefont {A.}~\bibnamefont
  {Javadi-Abhari}}, \bibinfo {author} {\bibfnamefont {L.~S.}\ \bibnamefont
  {Bishop}}, \bibinfo {author} {\bibfnamefont {I.}~\bibnamefont {Lauer}},
  \bibinfo {author} {\bibfnamefont {D.~F.}\ \bibnamefont {Bogorin}}, \bibinfo
  {author} {\bibfnamefont {M.}~\bibnamefont {Brink}}, \bibinfo {author}
  {\bibfnamefont {L.}~\bibnamefont {Capelluto}}, \bibinfo {author}
  {\bibfnamefont {O.}~\bibnamefont {Günlük}}, \bibinfo {author}
  {\bibfnamefont {T.}~\bibnamefont {Itoko}}, \bibinfo {author} {\bibfnamefont
  {N.}~\bibnamefont {Kanazawa}}, \bibinfo {author} {\bibfnamefont
  {A.}~\bibnamefont {Kandala}}, \bibinfo {author} {\bibfnamefont {G.~A.}\
  \bibnamefont {Keefe}}, \bibinfo {author} {\bibfnamefont {K.}~\bibnamefont
  {Krsulich}}, \bibinfo {author} {\bibfnamefont {W.}~\bibnamefont {Landers}},
  \bibinfo {author} {\bibfnamefont {E.~P.}\ \bibnamefont {Lewandowski}},
  \bibinfo {author} {\bibfnamefont {D.~T.}\ \bibnamefont {McClure}}, \bibinfo
  {author} {\bibfnamefont {G.}~\bibnamefont {Nannicini}}, \bibinfo {author}
  {\bibfnamefont {A.}~\bibnamefont {Narasgond}}, \bibinfo {author}
  {\bibfnamefont {H.~M.}\ \bibnamefont {Nayfeh}}, \bibinfo {author}
  {\bibfnamefont {E.}~\bibnamefont {Pritchett}}, \bibinfo {author}
  {\bibfnamefont {M.~B.}\ \bibnamefont {Rothwell}}, \bibinfo {author}
  {\bibfnamefont {S.}~\bibnamefont {Srinivasan}}, \bibinfo {author}
  {\bibfnamefont {N.}~\bibnamefont {Sundaresan}}, \bibinfo {author}
  {\bibfnamefont {C.}~\bibnamefont {Wang}}, \bibinfo {author} {\bibfnamefont
  {K.~X.}\ \bibnamefont {Wei}}, \bibinfo {author} {\bibfnamefont {C.~J.}\
  \bibnamefont {Wood}}, \bibinfo {author} {\bibfnamefont {J.-B.}\ \bibnamefont
  {Yau}}, \bibinfo {author} {\bibfnamefont {E.~J.}\ \bibnamefont {Zhang}},
  \bibinfo {author} {\bibfnamefont {O.~E.}\ \bibnamefont {Dial}}, \bibinfo
  {author} {\bibfnamefont {J.~M.}\ \bibnamefont {Chow}},\ and\ \bibinfo
  {author} {\bibfnamefont {J.~M.}\ \bibnamefont {Gambetta}},\ }\bibfield
  {title} {\bibinfo {title} {Demonstration of quantum volume 64 on a
  superconducting quantum computing system},\ }\href
  {https://doi.org/10.1088/2058-9565/abe519} {\bibfield  {journal} {\bibinfo
  {journal} {Quantum Science and Technology}\ }\textbf {\bibinfo {volume}
  {6}},\ \bibinfo {pages} {025020} (\bibinfo {year} {2021})}\BibitemShut
  {NoStop}%
\bibitem [{\citenamefont {Eddins}\ \emph {et~al.}(2021)\citenamefont {Eddins},
  \citenamefont {Motta}, \citenamefont {Gujarati}, \citenamefont {Bravyi},
  \citenamefont {Mezzacapo}, \citenamefont {Hadfield},\ and\ \citenamefont
  {Sheldon}}]{eddins2021doubling}%
  \BibitemOpen
  \bibfield  {author} {\bibinfo {author} {\bibfnamefont {A.}~\bibnamefont
  {Eddins}}, \bibinfo {author} {\bibfnamefont {M.}~\bibnamefont {Motta}},
  \bibinfo {author} {\bibfnamefont {T.~P.}\ \bibnamefont {Gujarati}}, \bibinfo
  {author} {\bibfnamefont {S.}~\bibnamefont {Bravyi}}, \bibinfo {author}
  {\bibfnamefont {A.}~\bibnamefont {Mezzacapo}}, \bibinfo {author}
  {\bibfnamefont {C.}~\bibnamefont {Hadfield}},\ and\ \bibinfo {author}
  {\bibfnamefont {S.}~\bibnamefont {Sheldon}},\ }\bibfield  {title} {\bibinfo
  {title} {Doubling the size of quantum simulators by entanglement forging},\
  }\href {https://arxiv.org/abs/2104.10220} {\  (\bibinfo {year} {2021})},\
  \Eprint {https://arxiv.org/abs/2104.10220} {arXiv:2104.10220 [quant-ph]}
  \BibitemShut {NoStop}%
\bibitem [{\citenamefont {Alsina}\ and\ \citenamefont
  {Latorre}(2016)}]{Alsina2016}%
  \BibitemOpen
  \bibfield  {author} {\bibinfo {author} {\bibfnamefont {D.}~\bibnamefont
  {Alsina}}\ and\ \bibinfo {author} {\bibfnamefont {J.~I.}\ \bibnamefont
  {Latorre}},\ }\bibfield  {title} {\bibinfo {title} {Experimental test of
  mermin inequalities on a five-qubit quantum computer},\ }\href
  {https://doi.org/10.1103/PhysRevA.94.012314} {\bibfield  {journal} {\bibinfo
  {journal} {Phys. Rev. A}\ }\textbf {\bibinfo {volume} {94}},\ \bibinfo
  {pages} {012314} (\bibinfo {year} {2016})}\BibitemShut {NoStop}%
\bibitem [{\citenamefont {Mooney}\ \emph {et~al.}(2019)\citenamefont {Mooney},
  \citenamefont {Hill},\ and\ \citenamefont
  {Hollenberg}}]{mooney2019entanglement}%
  \BibitemOpen
  \bibfield  {author} {\bibinfo {author} {\bibfnamefont {G.~J.}\ \bibnamefont
  {Mooney}}, \bibinfo {author} {\bibfnamefont {C.~D.}\ \bibnamefont {Hill}},\
  and\ \bibinfo {author} {\bibfnamefont {L.~C.}\ \bibnamefont {Hollenberg}},\
  }\bibfield  {title} {\bibinfo {title} {Entanglement in a 20-qubit
  superconducting quantum computer},\ }\href
  {https://www.nature.com/articles/s41598-019-49805-7} {\bibfield  {journal}
  {\bibinfo  {journal} {Scientific reports}\ }\textbf {\bibinfo {volume} {9}},\
  \bibinfo {pages} {1} (\bibinfo {year} {2019})}\BibitemShut {NoStop}%
\bibitem [{\citenamefont {Wang}\ \emph {et~al.}(2018)\citenamefont {Wang},
  \citenamefont {Li}, \citenamefont {Yin},\ and\ \citenamefont
  {Zeng}}]{wang201816}%
  \BibitemOpen
  \bibfield  {author} {\bibinfo {author} {\bibfnamefont {Y.}~\bibnamefont
  {Wang}}, \bibinfo {author} {\bibfnamefont {Y.}~\bibnamefont {Li}}, \bibinfo
  {author} {\bibfnamefont {Z.-q.}\ \bibnamefont {Yin}},\ and\ \bibinfo {author}
  {\bibfnamefont {B.}~\bibnamefont {Zeng}},\ }\bibfield  {title} {\bibinfo
  {title} {16-qubit ibm universal quantum computer can be fully entangled},\
  }\href {https://www.nature.com/articles/s41534-018-0095-x} {\bibfield
  {journal} {\bibinfo  {journal} {npj Quantum information}\ }\textbf {\bibinfo
  {volume} {4}},\ \bibinfo {pages} {1} (\bibinfo {year} {2018})}\BibitemShut
  {NoStop}%
\bibitem [{\citenamefont {Wilde}\ and\ \citenamefont
  {Mizel}(2012)}]{wilde2012addressing}%
  \BibitemOpen
  \bibfield  {author} {\bibinfo {author} {\bibfnamefont {M.~M.}\ \bibnamefont
  {Wilde}}\ and\ \bibinfo {author} {\bibfnamefont {A.}~\bibnamefont {Mizel}},\
  }\bibfield  {title} {\bibinfo {title} {Addressing the clumsiness loophole in
  a leggett-garg test of macrorealism},\ }\href
  {https://link.springer.com/article/10.1007/s10701-011-9598-4} {\bibfield
  {journal} {\bibinfo  {journal} {Foundations of Physics}\ }\textbf {\bibinfo
  {volume} {42}},\ \bibinfo {pages} {256} (\bibinfo {year} {2012})}\BibitemShut
  {NoStop}%
\bibitem [{\citenamefont {Brunner}\ \emph {et~al.}(2014)\citenamefont
  {Brunner}, \citenamefont {Cavalcanti}, \citenamefont {Pironio}, \citenamefont
  {Scarani},\ and\ \citenamefont {Wehner}}]{Brunner2014}%
  \BibitemOpen
  \bibfield  {author} {\bibinfo {author} {\bibfnamefont {N.}~\bibnamefont
  {Brunner}}, \bibinfo {author} {\bibfnamefont {D.}~\bibnamefont {Cavalcanti}},
  \bibinfo {author} {\bibfnamefont {S.}~\bibnamefont {Pironio}}, \bibinfo
  {author} {\bibfnamefont {V.}~\bibnamefont {Scarani}},\ and\ \bibinfo {author}
  {\bibfnamefont {S.}~\bibnamefont {Wehner}},\ }\bibfield  {title} {\bibinfo
  {title} {Bell nonlocality},\ }\href
  {https://doi.org/10.1103/RevModPhys.86.419} {\bibfield  {journal} {\bibinfo
  {journal} {Rev. Mod. Phys.}\ }\textbf {\bibinfo {volume} {86}},\ \bibinfo
  {pages} {419} (\bibinfo {year} {2014})}\BibitemShut {NoStop}%
\bibitem [{\citenamefont {Benenti}\ \emph {et~al.}()\citenamefont {Benenti},
  \citenamefont {Casati}, \citenamefont {Rossini},\ and\ \citenamefont
  {Strini}}]{benenti2019principles}%
  \BibitemOpen
  \bibfield  {author} {\bibinfo {author} {\bibfnamefont {G.}~\bibnamefont
  {Benenti}}, \bibinfo {author} {\bibfnamefont {G.}~\bibnamefont {Casati}},
  \bibinfo {author} {\bibfnamefont {D.}~\bibnamefont {Rossini}},\ and\ \bibinfo
  {author} {\bibfnamefont {G.}~\bibnamefont {Strini}},\ }\href@noop {} {\emph
  {\bibinfo {title} {Principles of Quantum Computation and Information: A
  Comprehensive Textbook}}}\ (\bibinfo  {publisher} {World
  Scientific})\BibitemShut {NoStop}%
\bibitem [{\citenamefont {Dressel}\ and\ \citenamefont
  {Korotkov}(2014)}]{Dressel2014}%
  \BibitemOpen
  \bibfield  {author} {\bibinfo {author} {\bibfnamefont {J.}~\bibnamefont
  {Dressel}}\ and\ \bibinfo {author} {\bibfnamefont {A.~N.}\ \bibnamefont
  {Korotkov}},\ }\bibfield  {title} {\bibinfo {title} {Avoiding loopholes with
  hybrid bell-leggett-garg inequalities},\ }\href
  {https://doi.org/10.1103/PhysRevA.89.012125} {\bibfield  {journal} {\bibinfo
  {journal} {Phys. Rev. A}\ }\textbf {\bibinfo {volume} {89}},\ \bibinfo
  {pages} {012125} (\bibinfo {year} {2014})}\BibitemShut {NoStop}%
\bibitem [{\citenamefont {White}\ \emph {et~al.}(2016)\citenamefont {White},
  \citenamefont {Mutus}, \citenamefont {Dressel}, \citenamefont {Kelly},
  \citenamefont {Barends}, \citenamefont {Jeffrey}, \citenamefont {Sank},
  \citenamefont {Megrant}, \citenamefont {Campbell}, \citenamefont {Chen} \emph
  {et~al.}}]{white2016preserving}%
  \BibitemOpen
  \bibfield  {author} {\bibinfo {author} {\bibfnamefont {T.~C.}\ \bibnamefont
  {White}}, \bibinfo {author} {\bibfnamefont {J.}~\bibnamefont {Mutus}},
  \bibinfo {author} {\bibfnamefont {J.}~\bibnamefont {Dressel}}, \bibinfo
  {author} {\bibfnamefont {J.}~\bibnamefont {Kelly}}, \bibinfo {author}
  {\bibfnamefont {R.}~\bibnamefont {Barends}}, \bibinfo {author} {\bibfnamefont
  {E.}~\bibnamefont {Jeffrey}}, \bibinfo {author} {\bibfnamefont
  {D.}~\bibnamefont {Sank}}, \bibinfo {author} {\bibfnamefont {A.}~\bibnamefont
  {Megrant}}, \bibinfo {author} {\bibfnamefont {B.}~\bibnamefont {Campbell}},
  \bibinfo {author} {\bibfnamefont {Y.}~\bibnamefont {Chen}}, \emph {et~al.},\
  }\bibfield  {title} {\bibinfo {title} {Preserving entanglement during weak
  measurement demonstrated with a violation of the bell--leggett--garg
  inequality},\ }\href {https://www.nature.com/articles/npjqi201522} {\bibfield
   {journal} {\bibinfo  {journal} {npj Quantum Information}\ }\textbf {\bibinfo
  {volume} {2}},\ \bibinfo {pages} {1} (\bibinfo {year} {2016})}\BibitemShut
  {NoStop}%
\bibitem [{\citenamefont {Thenabadu}\ \emph {et~al.}(2020)\citenamefont
  {Thenabadu}, \citenamefont {Cheng}, \citenamefont {Pham}, \citenamefont
  {Drummond}, \citenamefont {Rosales-Z\'arate},\ and\ \citenamefont
  {Reid}}]{Thenabadu2020}%
  \BibitemOpen
  \bibfield  {author} {\bibinfo {author} {\bibfnamefont {M.}~\bibnamefont
  {Thenabadu}}, \bibinfo {author} {\bibfnamefont {G.-L.}\ \bibnamefont
  {Cheng}}, \bibinfo {author} {\bibfnamefont {T.~L.~H.}\ \bibnamefont {Pham}},
  \bibinfo {author} {\bibfnamefont {L.~V.}\ \bibnamefont {Drummond}}, \bibinfo
  {author} {\bibfnamefont {L.}~\bibnamefont {Rosales-Z\'arate}},\ and\ \bibinfo
  {author} {\bibfnamefont {M.~D.}\ \bibnamefont {Reid}},\ }\bibfield  {title}
  {\bibinfo {title} {Testing macroscopic local realism using local nonlinear
  dynamics and time settings},\ }\href
  {https://doi.org/10.1103/PhysRevA.102.022202} {\bibfield  {journal} {\bibinfo
   {journal} {Phys. Rev. A}\ }\textbf {\bibinfo {volume} {102}},\ \bibinfo
  {pages} {022202} (\bibinfo {year} {2020})}\BibitemShut {NoStop}%
\bibitem [{\citenamefont {Thenabadu}\ and\ \citenamefont
  {Reid}(2021)}]{thenabadu2021bipartite}%
  \BibitemOpen
  \bibfield  {author} {\bibinfo {author} {\bibfnamefont {M.}~\bibnamefont
  {Thenabadu}}\ and\ \bibinfo {author} {\bibfnamefont {M.~D.}\ \bibnamefont
  {Reid}},\ }\bibfield  {title} {\bibinfo {title} {Bipartite leggett-garg and
  macroscopic bell inequality violations using cat states: distinguishing weak
  and deterministic macroscopic realism},\ }\href
  {https://journals.aps.org/pra/abstract/10.1103/PhysRevA.102.022202} {\
  (\bibinfo {year} {2021})},\ \Eprint {https://arxiv.org/abs/2012.14997}
  {arXiv:2012.14997 [quant-ph]} \BibitemShut {NoStop}%
\bibitem [{\citenamefont {Lieb}\ and\ \citenamefont
  {Robinson}(1972)}]{lieb1972finite}%
  \BibitemOpen
  \bibfield  {author} {\bibinfo {author} {\bibfnamefont {E.~H.}\ \bibnamefont
  {Lieb}}\ and\ \bibinfo {author} {\bibfnamefont {D.~W.}\ \bibnamefont
  {Robinson}},\ }\bibfield  {title} {\bibinfo {title} {The finite group
  velocity of quantum spin systems},\ }\bibfield  {booktitle} {\emph {\bibinfo
  {booktitle} {Statistical mechanics}},\ }\href
  {https://link.springer.com/article/10.1007/BF01645779} {\ ,\ \bibinfo {pages}
  {425} (\bibinfo {year} {1972})}\BibitemShut {NoStop}%
\bibitem [{\citenamefont {Emary}\ \emph {et~al.}(2013)\citenamefont {Emary},
  \citenamefont {Lambert},\ and\ \citenamefont {Nori}}]{emary2013leggett}%
  \BibitemOpen
  \bibfield  {author} {\bibinfo {author} {\bibfnamefont {C.}~\bibnamefont
  {Emary}}, \bibinfo {author} {\bibfnamefont {N.}~\bibnamefont {Lambert}},\
  and\ \bibinfo {author} {\bibfnamefont {F.}~\bibnamefont {Nori}},\ }\bibfield
  {title} {\bibinfo {title} {Leggett--garg inequalities},\ }\href
  {https://iopscience.iop.org/article/10.1088/0034-4885/77/1/016001} {\bibfield
   {journal} {\bibinfo  {journal} {Reports on Progress in Physics}\ }\textbf
  {\bibinfo {volume} {77}},\ \bibinfo {pages} {016001} (\bibinfo {year}
  {2013})}\BibitemShut {NoStop}%
\bibitem [{\citenamefont {Majidy}\ \emph {et~al.}(2019)\citenamefont {Majidy},
  \citenamefont {Katiyar}, \citenamefont {Anikeeva}, \citenamefont
  {Halliwell},\ and\ \citenamefont {Laflamme}}]{SSMajidy2019}%
  \BibitemOpen
  \bibfield  {author} {\bibinfo {author} {\bibfnamefont {S.-S.}\ \bibnamefont
  {Majidy}}, \bibinfo {author} {\bibfnamefont {H.}~\bibnamefont {Katiyar}},
  \bibinfo {author} {\bibfnamefont {G.}~\bibnamefont {Anikeeva}}, \bibinfo
  {author} {\bibfnamefont {J.}~\bibnamefont {Halliwell}},\ and\ \bibinfo
  {author} {\bibfnamefont {R.}~\bibnamefont {Laflamme}},\ }\bibfield  {title}
  {\bibinfo {title} {Exploration of an augmented set of leggett-garg
  inequalities using a noninvasive continuous-in-time velocity measurement},\
  }\href {https://doi.org/10.1103/PhysRevA.100.042325} {\bibfield  {journal}
  {\bibinfo  {journal} {Phys. Rev. A}\ }\textbf {\bibinfo {volume} {100}},\
  \bibinfo {pages} {042325} (\bibinfo {year} {2019})}\BibitemShut {NoStop}%
\bibitem [{\citenamefont {Majidy}\ \emph {et~al.}(2021)\citenamefont {Majidy},
  \citenamefont {Halliwell},\ and\ \citenamefont {Laflamme}}]{SSMajidy2021}%
  \BibitemOpen
  \bibfield  {author} {\bibinfo {author} {\bibfnamefont {S.}~\bibnamefont
  {Majidy}}, \bibinfo {author} {\bibfnamefont {J.~J.}\ \bibnamefont
  {Halliwell}},\ and\ \bibinfo {author} {\bibfnamefont {R.}~\bibnamefont
  {Laflamme}},\ }\bibfield  {title} {\bibinfo {title} {Detecting violations of
  macrorealism when the original leggett-garg inequalities are satisfied},\
  }\href {https://doi.org/10.1103/PhysRevA.103.062212} {\bibfield  {journal}
  {\bibinfo  {journal} {Phys. Rev. A}\ }\textbf {\bibinfo {volume} {103}},\
  \bibinfo {pages} {062212} (\bibinfo {year} {2021})}\BibitemShut {NoStop}%
\bibitem [{\citenamefont {Budroni}\ and\ \citenamefont
  {Emary}(2014)}]{Budroni2014}%
  \BibitemOpen
  \bibfield  {author} {\bibinfo {author} {\bibfnamefont {C.}~\bibnamefont
  {Budroni}}\ and\ \bibinfo {author} {\bibfnamefont {C.}~\bibnamefont
  {Emary}},\ }\bibfield  {title} {\bibinfo {title} {Temporal quantum
  correlations and leggett-garg inequalities in multilevel systems},\ }\href
  {https://doi.org/10.1103/PhysRevLett.113.050401} {\bibfield  {journal}
  {\bibinfo  {journal} {Phys. Rev. Lett.}\ }\textbf {\bibinfo {volume} {113}},\
  \bibinfo {pages} {050401} (\bibinfo {year} {2014})}\BibitemShut {NoStop}%
\bibitem [{\citenamefont {Lambert}\ \emph {et~al.}(2016)\citenamefont
  {Lambert}, \citenamefont {Debnath}, \citenamefont {Kockum}, \citenamefont
  {Knee}, \citenamefont {Munro},\ and\ \citenamefont {Nori}}]{Lambert2016}%
  \BibitemOpen
  \bibfield  {author} {\bibinfo {author} {\bibfnamefont {N.}~\bibnamefont
  {Lambert}}, \bibinfo {author} {\bibfnamefont {K.}~\bibnamefont {Debnath}},
  \bibinfo {author} {\bibfnamefont {A.~F.}\ \bibnamefont {Kockum}}, \bibinfo
  {author} {\bibfnamefont {G.~C.}\ \bibnamefont {Knee}}, \bibinfo {author}
  {\bibfnamefont {W.~J.}\ \bibnamefont {Munro}},\ and\ \bibinfo {author}
  {\bibfnamefont {F.}~\bibnamefont {Nori}},\ }\bibfield  {title} {\bibinfo
  {title} {Leggett-garg inequality violations with a large ensemble of
  qubits},\ }\href {https://doi.org/10.1103/PhysRevA.94.012105} {\bibfield
  {journal} {\bibinfo  {journal} {Phys. Rev. A}\ }\textbf {\bibinfo {volume}
  {94}},\ \bibinfo {pages} {012105} (\bibinfo {year} {2016})}\BibitemShut
  {NoStop}%
\bibitem [{\citenamefont {Fritz}(2010)}]{fritz2010quantum}%
  \BibitemOpen
  \bibfield  {author} {\bibinfo {author} {\bibfnamefont {T.}~\bibnamefont
  {Fritz}},\ }\bibfield  {title} {\bibinfo {title} {Quantum correlations in the
  temporal clauser--horne--shimony--holt (chsh) scenario},\ }\href
  {https://iopscience.iop.org/article/10.1088/1367-2630/12/8/083055} {\bibfield
   {journal} {\bibinfo  {journal} {New Journal of Physics}\ }\textbf {\bibinfo
  {volume} {12}},\ \bibinfo {pages} {083055} (\bibinfo {year}
  {2010})}\BibitemShut {NoStop}%
\bibitem [{\citenamefont {Huelga}\ \emph {et~al.}(1995)\citenamefont {Huelga},
  \citenamefont {Marshall},\ and\ \citenamefont {Santos}}]{huelga1995proposed}%
  \BibitemOpen
  \bibfield  {author} {\bibinfo {author} {\bibfnamefont {S.~F.}\ \bibnamefont
  {Huelga}}, \bibinfo {author} {\bibfnamefont {T.~W.}\ \bibnamefont
  {Marshall}},\ and\ \bibinfo {author} {\bibfnamefont {E.}~\bibnamefont
  {Santos}},\ }\bibfield  {title} {\bibinfo {title} {Proposed test for realist
  theories using rydberg atoms coupled to a high-q resonator},\ }\href
  {https://journals.aps.org/pra/abstract/10.1103/PhysRevA.52.R2497} {\bibfield
  {journal} {\bibinfo  {journal} {Physical Review A}\ }\textbf {\bibinfo
  {volume} {52}},\ \bibinfo {pages} {R2497} (\bibinfo {year}
  {1995})}\BibitemShut {NoStop}%
\bibitem [{\citenamefont {Cervera-Lierta}(2018)}]{cervera2018exact}%
  \BibitemOpen
  \bibfield  {author} {\bibinfo {author} {\bibfnamefont {A.}~\bibnamefont
  {Cervera-Lierta}},\ }\bibfield  {title} {\bibinfo {title} {Exact ising model
  simulation on a quantum computer},\ }\href
  {https://quantum-journal.org/papers/q-2018-12-21-114/} {\bibfield  {journal}
  {\bibinfo  {journal} {Quantum}\ }\textbf {\bibinfo {volume} {2}},\ \bibinfo
  {pages} {114} (\bibinfo {year} {2018})}\BibitemShut {NoStop}%
\bibitem [{\citenamefont {Smith}\ \emph {et~al.}(2019)\citenamefont {Smith},
  \citenamefont {Kim}, \citenamefont {Pollmann},\ and\ \citenamefont
  {Knolle}}]{smith2019simulating}%
  \BibitemOpen
  \bibfield  {author} {\bibinfo {author} {\bibfnamefont {A.}~\bibnamefont
  {Smith}}, \bibinfo {author} {\bibfnamefont {M.}~\bibnamefont {Kim}}, \bibinfo
  {author} {\bibfnamefont {F.}~\bibnamefont {Pollmann}},\ and\ \bibinfo
  {author} {\bibfnamefont {J.}~\bibnamefont {Knolle}},\ }\bibfield  {title}
  {\bibinfo {title} {Simulating quantum many-body dynamics on a current digital
  quantum computer},\ }\href
  {https://www.nature.com/articles/s41534-019-0217-0} {\bibfield  {journal}
  {\bibinfo  {journal} {npj Quantum Information}\ }\textbf {\bibinfo {volume}
  {5}},\ \bibinfo {pages} {1} (\bibinfo {year} {2019})}\BibitemShut {NoStop}%
\end{thebibliography}%

\end{document}